\documentclass[10pt,twocolumn, letterpaper]{article}

\pdfoutput=1
\usepackage{arxiv}
\usepackage{graphicx}
\usepackage{amssymb}
\usepackage{booktabs}
\usepackage{subcaption}
\usepackage{times}
\usepackage{nicefrac}
\usepackage{microtype}
\usepackage[left=2.35cm,right=2.35cm,top=2.7cm,bottom=2.7cm,a4paper]{geometry}

\newcommand{\eqnref}[1]{\mbox{Eqn~(\ref{#1})}}
\usepackage[pagebackref=true,breaklinks=true,letterpaper=true,colorlinks,bookmarks=false]{hyperref}

\iccvfinalcopy

\begin{document}

\title{Toward Interpretable Music Tagging with Self-Attention}

\author{
Minz Won$^1$\thanks{Correspondence to \tt minz.won@upf.edu}~~~~~~~~Sanghyuk Chun$^2$~~~~~~~~Xavier Serra$^1$
~\\
$^1$Music Technology Group, Universitat Pompeu Fabra, Barcelona, Spain \\
$^2$Clova AI Research, NAVER Corp. Seongnam, Korea
}

\maketitle

\begin{abstract}
Self-attention is an attention mechanism that learns a representation by relating different positions in the sequence. The transformer, which is a sequence model solely based on self-attention, and its variants achieved state-of-the-art results in many natural language processing tasks. Since music composes its semantics based on the relations between components in sparse positions, adopting the self-attention mechanism to solve music information retrieval (MIR) problems can be beneficial.

Hence, we propose a self-attention based deep sequence model for music tagging. 
The proposed architecture consists of shallow convolutional layers followed by stacked Transformer encoders. Compared to conventional approaches using fully convolutional or recurrent neural networks, our model is more interpretable while reporting competitive results. We validate the performance of our model with the MagnaTagATune and the Million Song Dataset. In addition, we demonstrate the interpretability of the proposed architecture with a heat map visualization.
\end{abstract}

\section{Introduction}\label{sec:introduction}
Following the huge successes in the fields of computer vision (CV) and natural language processing (NLP), convolutional neural networks (CNN) and recurrent neural networks (RNN) have successfully demonstrated their versatility in the field of music information retrieval (MIR). Deep architectures using CNN and RNN are now \textit{de facto} state-of-the-art in multiple MIR tasks including classification \cite{choi2017transfer,choi2016automatic, pons2017end,kim2018sample}, beat detection \cite{krebs2016downbeat}, music transcription \cite{vogl2017drum, hawthorne2017onsets}, and even music generation \cite{roberts2017hierarchical, huang2018musictransformer}. While traditional approaches in MIR extract relevant features for the target task based on domain knowledge, especially signal processing, recent works learn the features automatically from voluminous data using deep architectures.

Automatic music tagging is a multi-label classification task to predict music tags in accordance with the music audio contents. Music tags include high-level information, such as genre (rock, jazz), mood (happy, sad), and instrumentation (violin, guitar, piano), which can be utilized for music discovery and recommendation \cite{choi2016automatic}. Since CNN are powerful architectures that facilitate capturing local characteristics, their applications on music tagging could firmly establish the state-of-the-art results \cite{pons2017end, kim2018sample}.

However, we believe that music is sequential and it composes its high-level semantics based on the relations between individual components in long-term sparse positions, not only based on the local information. On analogous motivations, Choi et al. adopted convolutional recurrent neural networks (CRNN) \cite{choi2017convolutional} and Pons et al. tried to depict deep architectures in two parts: front-end and back-end \cite{pons2017end}. The front-end, which is equivalent to the CNN part of CRNN, learns local features. The back-end, which corresponds to the RNN part of CRNN, captures the structure of learned local features. Although they reported remarkable results, they are not suitable for modeling the long-term context. To encapsulate long-term context with CNN back-end, deep stacks of convolutional layers followed by subsampling layers (mostly max-pooling) are required, which will end up with blurred time resolution. RNN back-end with longer sequence inputs suffers from the demand of huge computational power and gradient vanishing/exploding problems \cite{pascanu2013difficulty}. In addition, CNN for MIR are yet less interpretable despite there has been noteworthy previous research to explain the predictions \cite{choi2016explaining, mishra2017local, mishraunderstanding}. One possible reason is that spectrogram-based 2D CNN models which have been used in the research learn spectro-temporal characteristics in each layer, while music is a temporal sequence of individual audio events.

Self-attention is an attention mechanism that learns a representation by relating different positions in the sequence. It facilitates the model to learn long-term context by relating each pair of positions directly. The transformer \cite{vaswani2017attention}, which is a sequence model solely based on self-attention, and its variants \cite{devlin2018bert,radford2018language} showed compelling results on extensive NLP tasks. Inspired by this, we propose to adopt the successful architecture to the back-end of music tagging models. By this means, one can expect not only the performance but also the interpretability.

In the following section (Section \ref{sec:related}), we review related music tagging models and the self-attention mechanism in detail. Then we depict the architecture of the proposed model (Section \ref{sec:model}) and dataset (Section \ref{sec:dataset}). 
Section \ref{sec:results} includes experimental results, careful ablation studies, and interpretable visualization of attention maps. 
Finally, we conclude with future works in Section \ref{sec:conclusion}.

\section{Related work}\label{sec:related}
\subsection{Deep Architectures for Music Tagging}
Choi et al. proposed to use fully convolutional networks (FCN) for music tagging \cite{choi2016automatic}. 
This architecture is also called \textit{vgg}-ish CNN because it uses stacks of $3\times3$ convolution filters as proposed in \cite{simonyan2014very}.
It was broadly used to solve MIR problems since one can take advantage of time-frequency invariance and its robustness to distortion. 

Pons et al. exploited domain knowledge to elaborate musically motivated convolution filter designs for music tagging \cite{pons2017end}. Vertical \cite{pons2017timbre} and Horizontal \cite{pons2017designing} filters were designed to capture timbral and temporal information, respectively, and combinations of both filters could achieve superb results in music tagging. This architecture will be used as one of our baselines that uses spectrogram inputs.

Lee et al. proposed a more end-to-end oriented architecture design which uses raw audio as its input, known as sample-level CNN \cite{lee2017sample}. There is no need for short-time Fourier transform (STFT) to get spectrograms in this architecture. Sample-level CNN could demonstrate their appropriacy in MIR tasks \cite{lee2017sample,kim2018sample} and it is known that the sample-level CNN show better results in bigger datasets \cite{pons2017end}. We use sample-level CNN as our another baseline that uses raw audio inputs. 

To interpret trained models, previous works \cite{choi2016explaining, mishra2017local, mishraunderstanding} elaborated to visualize and auralize learned information. However, current visualization and auralization are yet less interpretable. 
We assume the reason is due to the model architecture, which learns local spectro-temporal information (with stacks of $3\times3$ filters) instead of modeling the input as a sequence of individual audio events.

\subsection{Self-attention}\label{subsec:attention}
The self-attention mechanism has become a substitute for RNN to capture a long-range structure within sequential data.
Unlike RNN, a self-attention module computes the response at a location in a sequence by attending to all locations within the same sequence.
Recently, the Transformer \cite{vaswani2017attention} has shown by solely using self-attention modules without RNN, the model could achieve state-of-the-art performance in neural machine translation (NMT) task.
Similarly, self-attention has achieved successful classification performance with interpretability in video classification \cite{wang2018nonlocal,DBLP:journals/corr/abs-1810-13125} and text classification \cite{lin2017structured} tasks.

Self-attention is also used for generative models such as generative adversarial networks  (GAN) \cite{zhang2018sagan} and auto-regressive models \cite{parmar2018imagetransformer, huang2018musictransformer}.
In particular, the Music Transformer \cite{huang2018musictransformer} has shown that self-attention modules could model the long term dependency for musical representations using symbolic data, such as MIDI. And Wave2Midi2Wave \cite{hawthorne2018enabling} expanded the research toward raw audio by adopting the Onsets and Frames \cite{hawthorne2017onsets} to transcribe the raw audio (wave2midi), the Music Transformer \cite{huang2018musictransformer} to generate MIDI notes, and the Wavenet \cite{van2016wavenet} to generate raw audio from the MIDI notes(midi2wave).

Finally, self-attention also achieved great success in large scale pre-trained language models such as Google BERT \cite{devlin2018bert} and OpenAI GPT-2 \cite{radford2018language}.
\section{Proposed architecture}\label{sec:model}
In this section, we introduce the main motivation of proposed research and depict the details of front-ends and back-ends that we used. 

Convolutional recurrent neural networks (CRNN) were designed to capture local characteristics and their temporal representations using convolutional layers and following recurrent layers, respectively. Motivated from successful applications of CRNN in document classification \cite{tang2015document}, image classification \cite{zuo2015convolutional}, and music transcription \cite{sigtia2016end}, Choi et al. adopted CRNN to automatic music tagging \cite{choi2017convolutional}. 

In the same context, Pons et al. proposed to divide deep neural networks for MIR into two parts: front-end and back-end \cite{pons2017end}. Front-end maps input signal to a latent-space and back-end predicts the output based on the obtained representations from the front-end.

In summary, the aforementioned two models are both using CNN front-end but one uses RNN back-end \cite{choi2017convolutional} and another uses CNN back-end \cite{pons2017end}. By this mean, we can expect the front-end to capture local information: e.g., timbre, pitch, and chord; and the back-end to capture more structural information: e.g., rhythmic patterns, melodic contours, and chord progressions; based on the combination of the captured local components. As we explored in Section \ref{subsec:attention}, previous research has already proven the robustness of self-attention for long-term sequence modeling by stacking them. Hence, we propose a music tagging model consists of CNN front-end and self-attention back-end. From now, we call each model as `\textit{Frontend}\_\textit{Backend}': e.g., \textit{Spec}\_\textit{Att} means a model using spectrogram based front-end and our attention based back-end. Following subsections denote the architectures of front-end and back-end that we used.

\begin{figure}
 \centerline{
 \includegraphics[width=\linewidth]{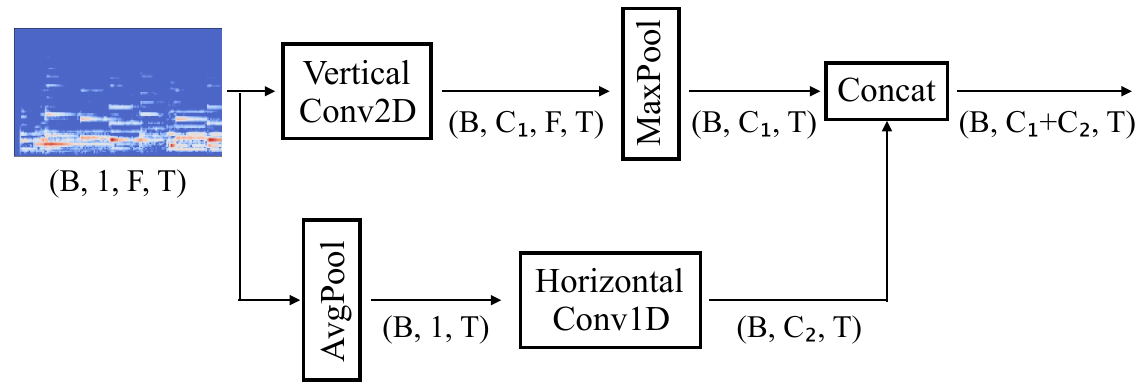}}
 \caption{\textit{Spec} front-end. B, C, F, and T stand for batch, channel, frequency, and time dimension.}
 \label{fig:spec}
\end{figure}

\begin{table}[]
\small
\begin{center}
\begin{tabular}{@{}cccc@{}}
\toprule
\multicolumn{2}{c}{Spec} & \multicolumn{2}{c}{Raw}  \\ \midrule
Layer             & Filter shape             & Layer                   & Filter shape    \\ \midrule
1   & $32\times38\times1$    &1  &$128\times3$ \\
1   & $32\times86\times1$    &2  &$128\times3$ \\
1   & $16\times38\times3$    &3  &$128\times3$ \\
1   & $16\times86\times3$    &4  &$256\times3$ \\
1   & $8\times38\times7$    &5  &$256\times3$ \\
1   & $8\times86\times7$    &6  &$256\times3$ \\
1   & $64\times33$    &7  &$256\times3$ \\
1   & $32\times65$    &8  &$256\times3$ \\
1   & $16\times129$    &9  &$256\times3$ \\
1   & $8\times165$    &10  &$512\times3$ \\ \bottomrule
\end{tabular}
\end{center}
\caption{Filter shapes of \textit{Spec} front-end and \textit{Raw} front/back-end. 
Dimensions of filters are $Channel \times Frequency \times Time$ or $Channel \times Time$.}
\label{tab:parameters}
\end{table}

\subsection{Front-end}
In accordance with previous work \cite{pons2017end}, two different front-ends were tested: 2D CNN using spectrogram inputs and 1D CNN using raw audio waveform. 

The 2D CNN front-end in our experiment is an architecture that can leverage domain knowledge \cite{pons2017end}. To facilitate learning timbral and temporal patterns, vertical and horizontal filter shapes were designed, respectively. Vertical filters \cite{pons2017timbre} capture short-time spectro-temporal features.
After the convolution on input spectrograms, extracted feature maps are max-pooled along the frequency axis. By this mean, the appearance of each instrument will be captured while pitch related information to be ignored. Horizontal filters \cite{pons2017designing} capture temporal energy flux patterns in up to 2.6s sequence.
Horizontal filters receive average-pooled (along with frequency axis) spectrograms as their inputs. Since vertical filters have a max-pooling layer after the convolutional layer, and horizontal filters have an average-pooling layer before the convolutional layer, the frequency axis of the tensors can be flattened --- see Figure \ref{fig:spec}. Flattened two feature maps are concatenated along channels. We call this spectrogram based front-end as \textit{Spec} front-end. 
\textit{Spec} front-end uses 256 frames ($\approx$4.1s) of spectrogram chunk as its input.

Sample-level CNN \cite{lee2017sample, kim2018sample} stack short grain of one dimensional convolution filters (e.g. $1\times3$) to model the music sequence. We call the front-end using sample-level CNN as \textit{Raw}. Strictly, there is no clear boundary of front-end and back-end in the sample-level CNN since it consists of homogeneous 1D convolutional layers. However, to examine our self-attention back-end, we regarded the first five convolutional layers as a front-end since one frame in the feature map after the five layers can include 15.2ms of audio which can be compared with one frame of spectrograms (16ms). Only when we use self-attention back-end, for the fair comparison, \textit{Raw} front-end is followed by one $1\times7$ convolutional layer since vertical filters of \textit{Spec} front-end have capacities of up to 7 frames (112ms). 
\textit{Raw} front-end uses 65,610 samples ($\approx$4.1s) of raw audio chunk.
Detailed number of parameters are described in Table \ref{tab:parameters}.

\begin{figure}
 \centerline{
 \includegraphics[width=\linewidth]{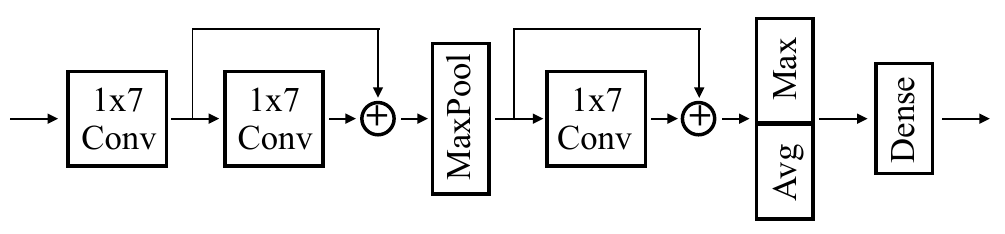}}
 \caption{\textit{CNN$_P$} back-end.}
 \label{fig:cnnp}
\end{figure}

\subsection{Back-end (CNN)}
The spectrogram back-end uses stacks of 1D CNN with residual connections \cite{he2016deep} --- see Figure \ref{fig:cnnp}. Channel size is 512 for each layer. We denote this back-end as \textit{CNN$_P$} name after Pons et al. \cite{pons2017end}. 

As we reviewed in the previous subsection, sample-level CNN do not have a clear boundary of front-end and back-end. For convenience, we call the latter five layers as \textit{CNN$_L$} back-end name after Lee et al. \cite{lee2017sample}. In the end, \textit{Raw}\_\textit{CNN$_L$} model consists of ten 1D convolutional layers as proposed in the original paper \cite{lee2017sample}. Each layer of both back-ends also uses batch normalization and ReLU non-linearity.

\subsection{Back-end (Self-Attention)}
\begin{figure}
 \centerline{
 \includegraphics[width=\linewidth]{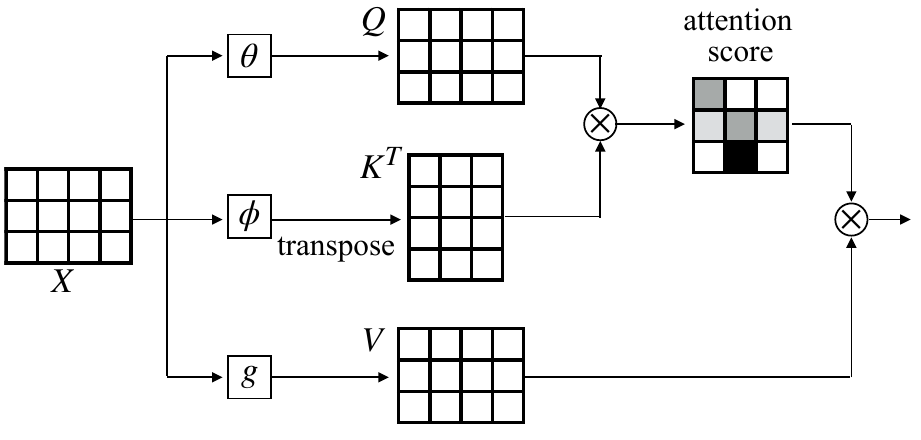}}
 \caption{Self-attention mechanism.}
 \label{fig:sa}
\end{figure}

In a field of NLP, self-attention is used to build higher-level semantic by relating each component appeared in a sequence. From the given query ($Q$), the machine learns the relation between the query and keys ($K$) to compute attention scores, and multiply the attention scores to the values ($V$). Finally, the sum of attended values composes the semantics of the given query. 
For example, there is a sentence ``I play bass". With ``bass" alone, we don't know if it is a fish or an instrument. We know it is an instrument based on the context because it has ``play" in the sentence. When we want to know the semantic of ``bass" ($Q$), we calculate the attention score by comparing the distance between ``bass" and other words ($K$) in the sequence: ``I", ``play", and ``bass". In this context, for given query ``bass", ``play" will have higher attention score than ``I" since ``play" is a more important component to make ``bass" as an instrument. As a result, the third frame of the output feature map, which is a position of ``bass", can have the context that the ``bass" is a musical instrument. 
Since the attention score is computed from the sentence itself, it is called self-attention or intra-attention. 
The Transformer \cite{vaswani2017attention}, which is a deep stack of self-attention, uses scaled dot-product attention to compute attention scores. This can be simply described as matrix multiplications:
\begin{equation}\label{eqn:attention}
Attention(Q, K, V) = softmax\left(\frac{QK^T}{\sqrt{d_k}}\right)V
\end{equation}
where $d_k$ is a dimension of keys and $Q$, $K$, $V$ are matrices whose shapes are $Sequence\times Embedding$.

We applied the self-attention to the feature map that we get from the front-end convolution. Suppose a convolution feature map is given after the front-end convolution of \textit{Spec} or \textit{Raw} and let \(X\in \mathbb{R}^{T\times CF}\) or \(X\in \mathbb{R}^{T\times C}\) denote the feature map, where $C$ is channels, $T$ is time, and $F$ is frequency axis. For simplification, here we only explain with \(X\in \mathbb{R}^{T\times C}\) which is a feature map of \textit{Raw} front-end. In this case, an $1\times C$ vector of the feature map at each time bin can be regarded as a word embedding. Hence, $Q$, $K$, and $V$ of the feature map $X$ can be denoted as:

\begin{equation}\label{eqn:qkv}
\begin{array}{l}
Q = \theta(X) = XW_\theta \in \mathbb{R}^{T\times C}\\
K = \phi(X) = XW_\phi \in \mathbb{R}^{T\times C}\\
V = g(X) = XW_g \in \mathbb{R}^{T\times C}
\end{array}
\end{equation}
where \(\theta(\cdot)\), \(\phi(\cdot)\), \(g(\cdot)\) are learnable transformations. 
If we omit softmax and the scaling factor from the \eqnref{eqn:attention} and apply \eqnref{eqn:qkv}:
\begin{equation}\label{eqn:ultimate}
Attention(X) = XW_\theta W_\phi^T X^T XW_g,
\end{equation}
which is a simple matrix multiplication form.
Figure \ref{fig:sa} shows a single self-attention layer that we described.

\begin{figure}
 \centerline{
 \includegraphics[width=\linewidth]{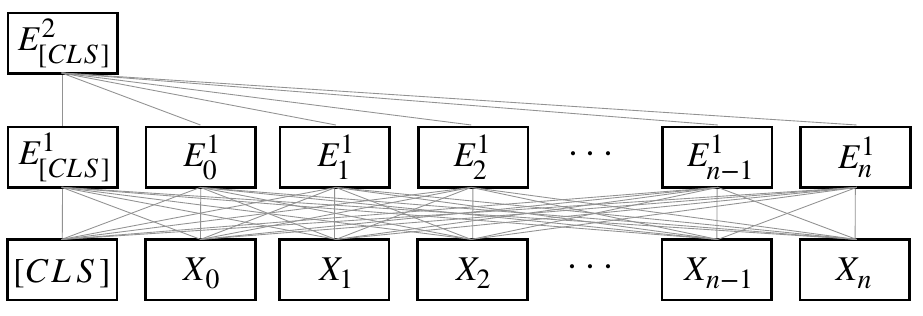}}
 \caption{\textit{Att} back-end with two self-attention layers.}
 \label{fig:bert}
\end{figure}

While the Transformer \cite{vaswani2017attention} has encoder and decoder parts to tackle the machine translation task, the Bidirectional Encoder Representations from
Transformers (BERT) \cite{devlin2018bert} only used encoders of the Transformer. Since our task is to classify, not to generate, we also only adopted the encoder part as BERT did. As shown in Figure \ref{fig:bert}, our proposed back-end uses stacks of self-attention to classify the tags of given sequence $X$. 
[$CLS$] is a special token that includes overall context for the classification.
We call our back-end
as \textit{Att} back-end to connote \textit{Attention}. Self-attention that we used is multi-head attention \cite{vaswani2017attention}.

Through this section, we described two front-ends: \textit{Spec} and \textit{Raw}; and three back-ends: \textit{CNN$_P$}, \textit{CNN$_L$}, and \textit{Att}. 
We set \textit{Spec}\_\textit{CNN$_P$} and \textit{Raw}\_\textit{CNN$_L$} as our baseline, 
which are equivalent to the original implementation of \cite{pons2017end} and \cite{lee2017sample}, respectively.
Then we experimented our back-end with \textit{Spec}\_\textit{Att} and \textit{Raw}\_\textit{Att} models. 

\subsection{Optimization}\label{subsec:optimization}
Careful design of learning rate schedule is critical to both of convergence speed and generalization \cite{hoffer2017train, seong2018towards}. 
ADAM \cite{kingma2014adam}, an adaptive optimization method, 
achieves fast convergence but it is generally known to impede the generalization of models \cite{keskar2017improving, wilson2017marginal}.
Instead of using conventional stochastic gradient descent (SGD) or ADAM, we propose an optimization technique inspired by the Switches from Adam to SGD (SWATS) \cite{keskar2017improving}.

We first optimize the network using ADAM \cite{kingma2014adam} with learning rate $1\mathrm{e}{-4}$, beta1 0.9, and beta2 0.999. After 60 epochs, we reload the model which achieved the best validation AUROC during the 60 epochs, and switch the optimizer to SGD with 
momentum 0.9 and nesterov momentum. 
We drop the learning rate by 10\% at the epoch 80 and 100.
In Section~\ref{subsec:ablation}, we show that our proposed mixed optimization scheme improves the generalization capacity than an SGD with manual learning rate scheduling.
Note that our proposed method loads the best model weights during the training while SWATS \cite{keskar2017improving} switches optimizer without changing the weights.

\section{Dataset}\label{sec:dataset}

\subsection{MagnaTagATune}
The MagnaTagATune (MTAT) dataset \cite{law2009evaluation} consists of $\approx$26k annotated audio clips with $\approx$30s duration. We only used top-50 tags as proposed in the previous work \cite{choi2016automatic} and followed the same data split of other research \cite{choi2016automatic, pons2017end, lee2017sample, kim2018sample}. Although the aforementioned works share the same data split, two recent works \cite{lee2017sample, kim2018sample} only used a refined subset. They removed songs that do not contain any of top-50 tags and $\approx$21k songs remained instead of $\approx$26k. Since this subset is more informative, we also used this in our experiments.

\subsection{Million Song Dataset}
For scalable research, we also explored a subset of the Million Song Dataset (MSD) with \texttt{Last.FM} tags \cite{bertin2011million}. Again, top-50 tags were selected \cite{choi2016automatic} and audio clips shorter than 29.1s were discarded. As a result, $\approx$242k songs were available. We followed the data split of \cite{lee2017sample}, \cite{kim2018sample} and \cite{pons2017end}.

\subsection{Preprocessing}
We investigate two different types of input: raw audio and log mel-spectrogram. 
For the comparable research, we decided to use 16kHz sampling rate for both inputs. Essentia library \cite{bogdanov2013essentia} was used to load and downsample the audio.

To get the log mel-spectrograms, hanning window of 512 samples with 50\% overlap has been used and the number of mel-bins was set to 96. Librosa library \cite{mcfee2015librosa} was used for this step. We did not normalize the dataset. Instead, \textit{CNN$_P$} has batch normalization in the first layer. 
\section{Results}\label{sec:results}

\begin{table}[]
\renewcommand{\tabcolsep}{0.8mm}
\small
\begin{tabular}{@{}cccccc@{}}
\toprule
& & \multicolumn{2}{c}{MTAT} & \multicolumn{2}{c}{MSD}  \\ \midrule
Front-end             & Back-end             & AUROC                   & AUPR                 & AUROC & AUPR \\ \midrule
\textit{Raw} \cite{lee2017sample} & \textit{CNN$_L$} \cite{lee2017sample} & 90.62                   & 44.20                & 88.42$^{*}$   & -  \\
\textit{Raw} \cite{lee2017sample} & \textit{Att} (Ours) & 90.66                   & 44.21                & 88.07   & 29.90  \\ \midrule
\textit{Spec} \cite{pons2017end} & \textit{CNN$_P$} \cite{pons2017end}  & 90.89                   & 45.03                & 88.75$^{*}$   & 31.24$^{*}$  \\
\textit{Spec} \cite{pons2017end} & \textit{Att} (Ours)                 & 90.80                   & 44.39                & 88.14   & 30.47  \\ \bottomrule
\end{tabular}
\caption{Comparison of state-of-art music tagging models on MTAT and MSD.
The results marked with (\mbox{*}) on top are reported values from the reference papers.}
\label{tab:main-results}
\end{table}

\begin{table}[]
\centering
\small
\begin{tabular}{@{}cccc@{}}
\toprule
\# heads & \# layers & AUROC & AUPR  \\ \midrule
1                 & 2                  & 87.73 & 36.93 \\
2                 & 2                  & 89.40 & 41.20 \\
3                 & 2                  & 90.23 & 43.23 \\
4                 & 2                  & 90.40 & 43.89 \\
5                 & 2                  & 90.60 & 43.91 \\
6                 & 2                  & 90.61 & 44.39 \\
7                 & 2                  & 90.74 & \textbf{44.43} \\
8                 & 2                  & \textbf{90.80} & 44.39 \\ \midrule
8                 & 1                  & 90.54 & 44.12 \\
8                 & 2                  & \textbf{90.80} & \textbf{44.39} \\
8                 & 3                  & 90.19 & 43.22 \\ \bottomrule
\end{tabular}
\caption{Impact of the number of attention heads and layers on MTAT.}
\label{tab:ablation-layers}
\end{table}

\subsection{Quantitative Results}
Following previous research \cite{pons2017end}, we report the Area Under Precision Recall curve (AUPR) along with conventional Area Under Receiver Operating Characteristic curve (AUROC). AUPR is known to be more informative to evaluate the algorithm's performance when it deals with highly skewed datasets \cite{davis2006relationship}. Since we are using user-generated tags (MTAT and MSD), there is popularity biased skewness in their distributions. Although we are using AUROC to choose the best model, it's not always the best in both metrics --- see Table \ref{tab:ablation-layers}.

Table \ref{tab:main-results} shows AUROC and AUPR of the baseline models and our proposed models. Each value in the table is the average of three different runs.
As shown in the table, our proposed \textit{Att} back-end reports competitive results for both datasets.

\begin{figure}[ht!]
    \centering
    \begin{subfigure}[h]{0.49\linewidth}
        \centering
        \includegraphics[width=\linewidth]{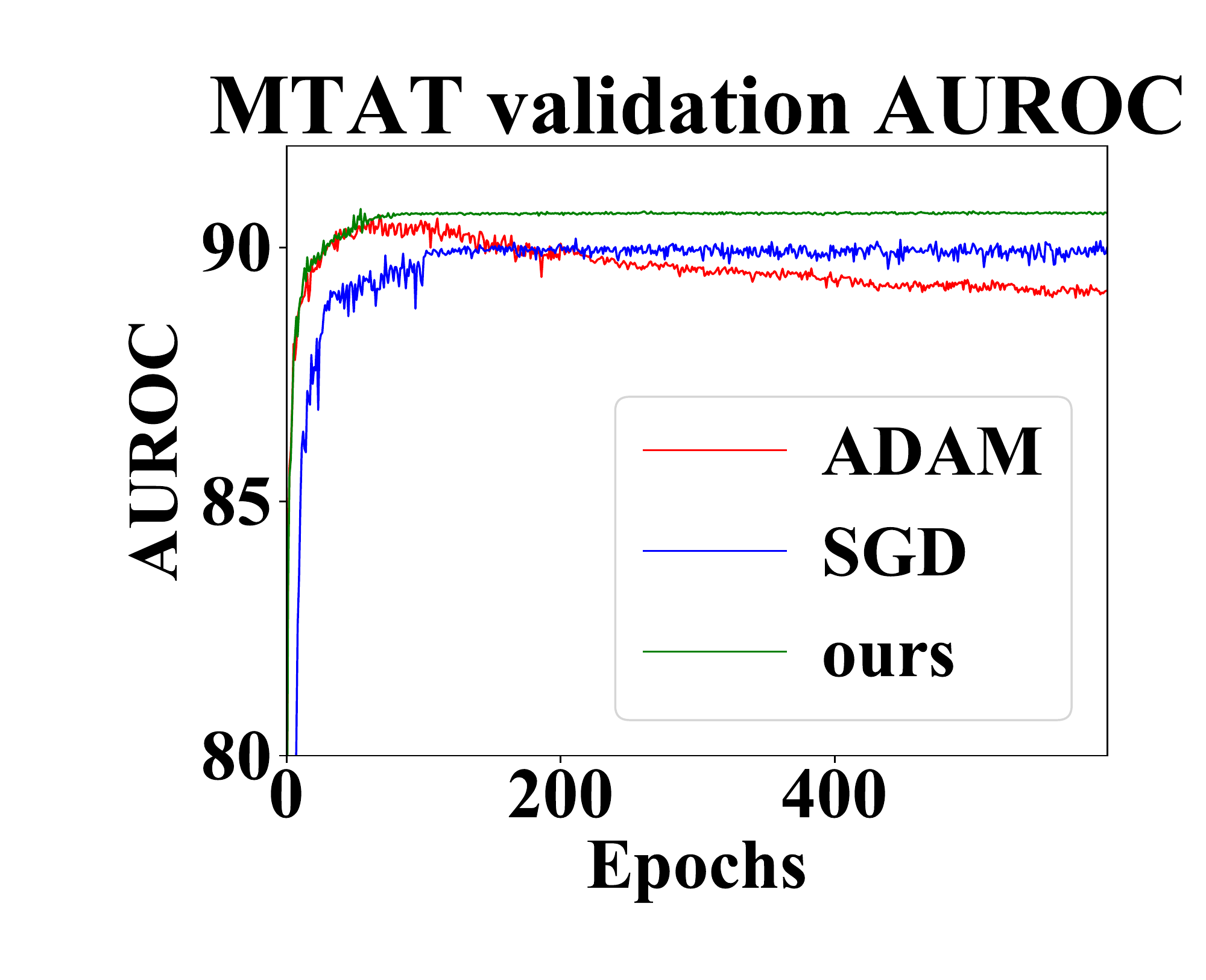}
    \end{subfigure}
    \begin{subfigure}[h]{0.49\linewidth}
        \centering
        \includegraphics[width=\linewidth]{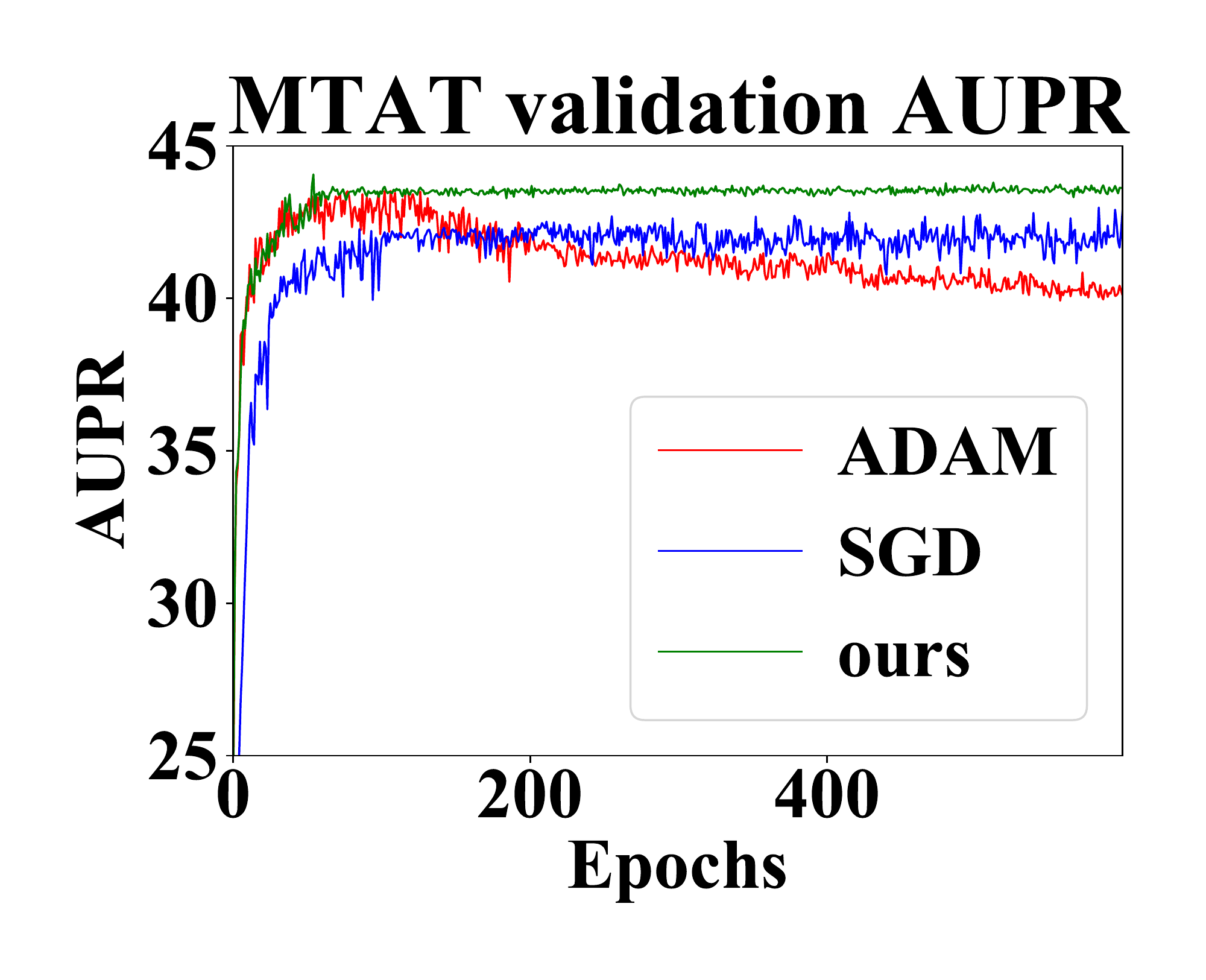}
    \end{subfigure}
    \caption{Comparison of optimizers: ADAM, SGD, and our proposed method.}
    \label{fig:opt_comp}
\end{figure}

\begin{table}[]
\centering
\small
\begin{tabular}{@{}cccc@{}}
\toprule
Input length & \# layers & AUROC & AUPR  \\ \midrule
256          & 2                 & 90.80 & 44.39 \\ \midrule
1024         & 2                 & 89.62 & 41.61 \\
1024         & 3                 & 89.85 & \textbf{42.25} \\
1024         & 4                 & \textbf{89.86} & 41.84 \\ \bottomrule
\end{tabular}
\caption{AUROC and AUPR results on MTAT using proposed \textit{Spec}\_\textit{Att} models with longer input sequence.}
\label{tab:long_input}
\end{table}

\subsection{Ablation Study}
\label{subsec:ablation}
    \noindent\textbf{Attention Parameters.} Choosing an appropriate number of attention layers and heads can be crucial for designing better models. As shown in Table \ref{tab:ablation-layers}, attention layers more than 2 did not show significant improvement and 8 attention heads reported the best performance. Hence, we fixed the number of attention layers and attention heads in our experiments as 2 and 8, respectively. Note that this setup is optimized for $\approx$4.1s inputs.

\noindent\textbf{Optimization.}
As we depicted in Section \ref{subsec:optimization}, we used our novel optimization method. By adopting ADAM \cite{kingma2014adam} in the beginning, we expected faster convergence than SGD. As shown in Figure \ref{fig:opt_comp}, ADAM and our optimization method show a steeper learning curve than SGD. However, AUROC and AUPR of ADAM go down after around 100 epochs, which means it failed to generalize the model. Since we switch our model to SGD at 60 epochs, it shows more stable learning curve than ADAM only. Although this switch point is an arbitrary point, our optimization method can generalize the model well because we load the best model during the training when we switch the optimizer or learning rate --- we used AUROC to choose the best model.

\noindent\textbf{Longer Sequence.}
In our main experiment, we only used relatively short audio chunks ($\approx$4.1s) as our input for the fair comparison --- sample-level CNN used short chunks. However, as we explained in Section \ref{subsec:attention}, self-attention is known to be efficient to model long-term sequence. We experimented the \textit{Spec}\_\textit{Att} model for MTAT using 1024 samples ($\approx$16.4s) and we could see slightly lower but comparable results --- see Table \ref{tab:long_input}. More stacks of self-attention layers were required to model longer sequence. 

\begin{figure*}[ht!]
    \centering
    \begin{subfigure}[h]{0.33\linewidth}
        \centering
        \includegraphics[width=\linewidth]{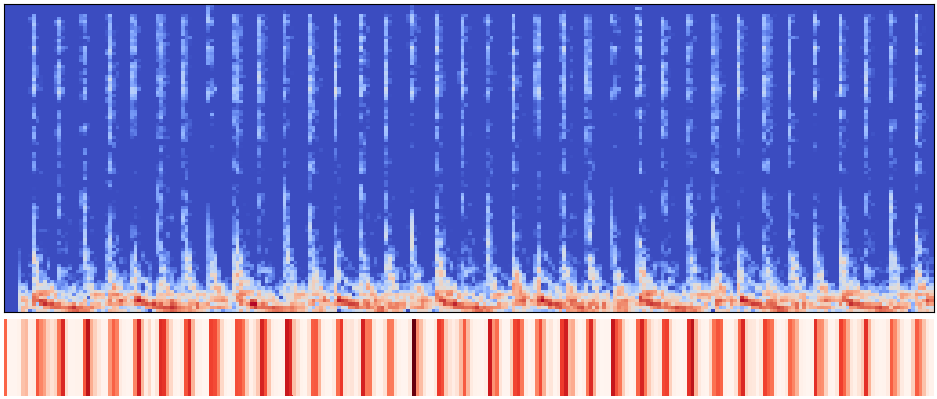}
        \caption{Tag - Beats}
        \label{fig:attention_heatmap_a}
    \end{subfigure}
    \begin{subfigure}[h]{0.33\linewidth}
        \centering
        \includegraphics[width=\linewidth]{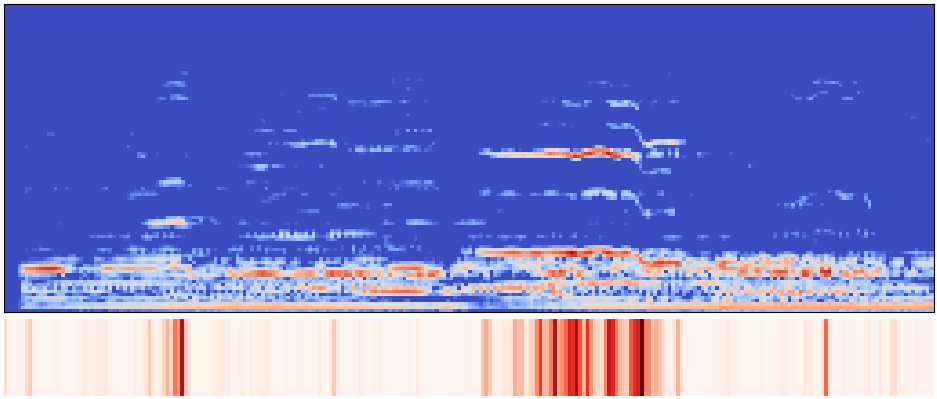}
        \caption{Tag - Female}
        \label{fig:attention_heatmap_b}
    \end{subfigure}
    \begin{subfigure}[h]{0.33\linewidth}
        \centering
        \includegraphics[width=\linewidth]{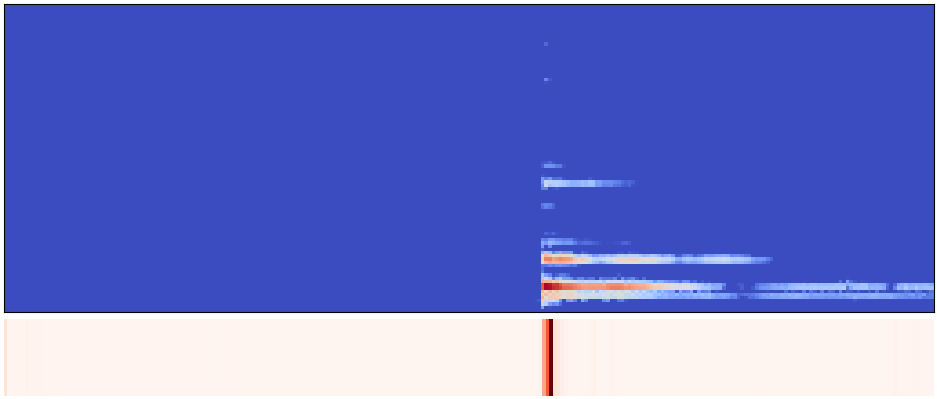}
        \caption{Tag - Quiet}
        \label{fig:attention_heatmap_c}
    \end{subfigure}
    \caption{Attention heat maps. More results are illustrated in the appendix.}
    \label{fig:attention_heatmap}
\end{figure*}

\begin{figure*}[ht!]
    \centering
    \begin{subfigure}[h]{0.33\linewidth}
        \centering
        \includegraphics[width=\linewidth]{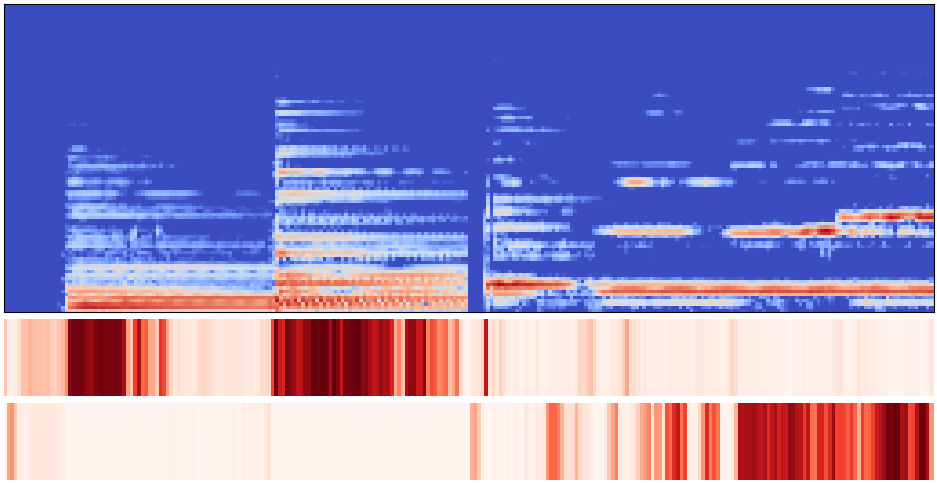}
        \caption{Piano + Flute}
        \label{fig:tagwise_contrib_a}
    \end{subfigure}
    \begin{subfigure}[h]{0.33\linewidth}
        \centering
        \includegraphics[width=\linewidth]{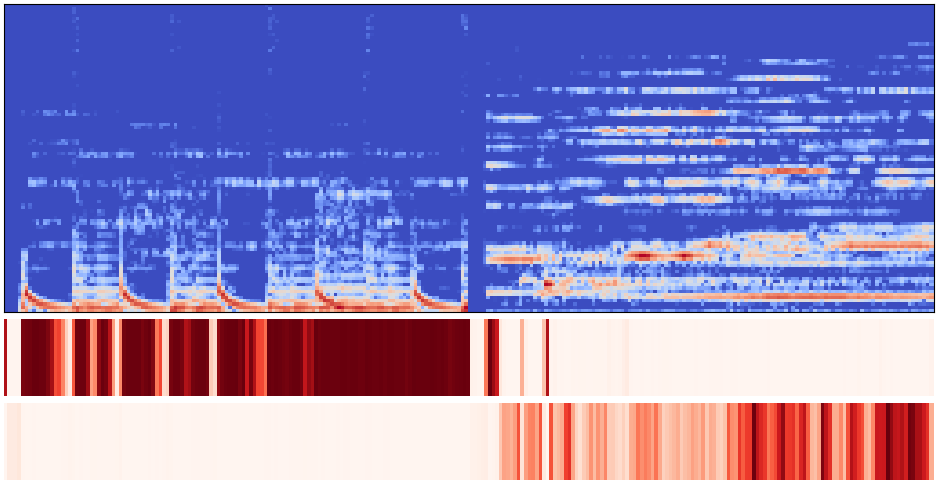}
        \caption{Techno + Classic}
        \label{fig:tagwise_contrib_b}
    \end{subfigure}
    \begin{subfigure}[h]{0.33\linewidth}
        \centering
        \includegraphics[width=\linewidth]{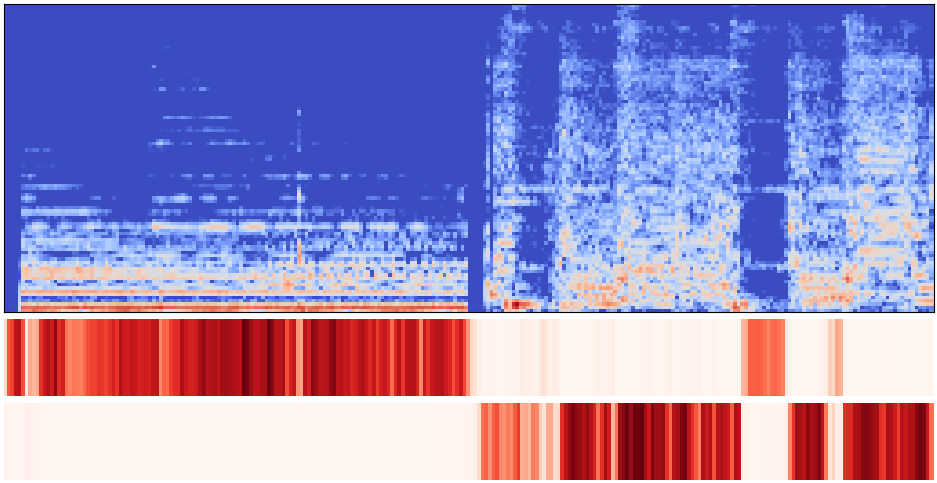}
        \caption{Quiet + Loud}
        \label{fig:tagwise_contrib_c}
    \end{subfigure}
    \caption{Tag-wise contribution heat maps on concatenated spectrograms. From the top, concatenated spectrograms, contribution heat maps to the first tags (Piano, Techno, and Quiet, respectively), and contribution heat maps to the second tags (Flute, Classic, and Loud, respectively). We report more results in the appendix.}
    \label{fig:tagwise_contrib}
\end{figure*}

Although self-attention is a powerful mechanism to model long sequential data, the amount of required memory increases quadratically by the sequence gets longer because we use dot-product attention. 

In order to secure bigger size of receptive field, the Image Transformer \cite{parmar2018imagetransformer} used local self-attention. The Compact Generalized Non-Local Network (CGNL) \cite{DBLP:journals/corr/abs-1810-13125} approximated the calculation of self-attention via a trilinear equation with Taylor expansion. To capture longer-term context from music audio, we can utilize these techniques to reduce the complexity of the model efficiently.

\subsection{Visualization}

To interpret the proposed model, we provide two different visualization: attention heat map and tag-wise contribution heat map. While attention heat map shows where the trained model pays more attention, tag-wise contribution heat map highlights which part of the input spectrogram is more relevant to predict the given tag.

\noindent\textbf{Attention Heat Map.}
To understand the behavior of the model, 
it is important to know which part of the audio the machine pays more attention to. To this end, we summed up attention scores from each attention head and visualized. Attention score $A$ of a single attention head can be described as:
\vspace{-0.5em}
\begin{equation}
A = softmax\left(\frac{QK^T}{\sqrt{d_k}}\right).
\end{equation}

Figure \ref{fig:attention_heatmap} shows log mel-spectrograms and according attention heat maps. For simplification, we only visualized the attention heat map of the last attention layer.
As we can see in Figure \ref{fig:attention_heatmap_a} and Figure \ref{fig:attention_heatmap_b}, the model pays more attention to relevant parts of spectrograms. However, we discovered one interesting thing which is: the model always pays attention to the parts with audio events. For example, in Figure \ref{fig:attention_heatmap_c}, the model pays attention to the loud part of the audio although the given spectrogram was classified as ``quiet". We could also observe this behavior from negative tags such as ``no vocal", ``no vocals",  and ``no voice".
One possible reason is that the model pays attention to the more informative part of the spectrogram. 
Indeed, negative tags report relatively worse AUROC ($\approx0.7$) than other tags ($\approx0.9$). 
Although attention heat maps can pinpoint where the machine pays attention for the decision, they cannot provide reasons for the classification or tagging.

\noindent\textbf{Tag-wise Contribution Heat Map.}
Understanding which part of the audio is more relevant to each tag is also important to interpret the model. We manually changed the attention score of the last attention layer. For each time step, we manipulated the attention score as 1 and set other parts as 0 so that we can see the contribution of each time bin to each tag. This tag-wise contribution heat map is inspired by the manual attention weight adjustment proposed in \cite{lee2017interactive}. To compare the different contribution of different audio, we concatenated two spectrograms and fed them through the network. For instance, Figure \ref{fig:tagwise_contrib_a} is a concatenated spectrogram of piano (left half) and flute (right half). The first row heat map highlights the contribution of each time bin to the ``piano" and the second row is for ``flute".
We repeated this for genre (Figure \ref{fig:tagwise_contrib_b}) and mood (Figure \ref{fig:tagwise_contrib_c}). 
As shown in Figure \ref{fig:tagwise_contrib_c}, the tag-wise contribution heat map can provide more information about tag specific part of the audio, which was not able to be observed from the attention heat map (Figure \ref{fig:attention_heatmap_c}).

\section{Conclusion}\label{sec:conclusion}

In this paper, we proposed a novel deep sequence model for music tagging which can facilitate better interpretability. The proposed model consists of CNN front-end and self-attention back-end. Experiments on MTAT dataset and MSD reported competitive results and we could demonstrate the interpretability of the model by visualizing attention heat maps and tag-wise contribution heat maps. By leveraging the acquired interpretation, one can obtain better intuition for the model design. Since proposed architecture is not task specific, it is expandable toward broad MIR tasks such as beat detection, rhythm classification, or music transcription.

\section{Acknowledgement}

This work was funded by the predoctoral grant MDM-2015-0502-17-2 from the Spanish Ministry of Economy and Competitiveness linked to the Maria de Maeztu Units of Excellence Programme (MDM-2015-0502).
Also, we acknowledge that the experiments were carried out on NAVER Smart Machine Learening (NSML) GPU platform \cite{nsml1, nsml2}.

{\small
\bibliographystyle{ieee}
\bibliography{references}

\begin{thebibliography}{10}\itemsep=-1pt

\bibitem{bertin2011million}
T.~Bertin-Mahieux, D.~P. Ellis, B.~Whitman, and P.~Lamere.
\newblock The million song dataset.
\newblock 2011.

\bibitem{bogdanov2013essentia}
D.~Bogdanov, N.~Wack, E.~G{\'o}mez~Guti{\'e}rrez, S.~Gulati, P.~Herrera~Boyer,
  O.~Mayor, G.~Roma~Trepat, J.~Salamon, J.~R. Zapata~Gonz{\'a}lez, and
  X.~Serra.
\newblock Essentia: An audio analysis library for music information retrieval.
\newblock In {\em Proceedings of the International Society for Music
  Information Retrieval Conference (ISMIR)}, 2013.

\bibitem{choi2016automatic}
K.~Choi, G.~Fazekas, and M.~Sandler.
\newblock Automatic tagging using deep convolutional neural networks.
\newblock {\em Proceedings of the International Society for Music Information
  Retrieval Conference (ISMIR)}, 2016.

\bibitem{choi2016explaining}
K.~Choi, G.~Fazekas, and M.~Sandler.
\newblock Explaining deep convolutional neural networks on music
  classification.
\newblock {\em arXiv preprint arXiv:1607.02444}, 2016.

\bibitem{choi2017convolutional}
K.~Choi, G.~Fazekas, M.~Sandler, and K.~Cho.
\newblock Convolutional recurrent neural networks for music classification.
\newblock In {\em Proceedings of the IEEE International Conference on
  Acoustics, Speech and Signal Processing (ICASSP)}, 2017.

\bibitem{choi2017transfer}
K.~Choi, G.~Fazekas, M.~Sandler, and K.~Cho.
\newblock Transfer learning for music classification and regression tasks.
\newblock {\em Proceedings of the International Society of Music Information
  Retrieval Conference (ISMIR)}, 2017.

\bibitem{davis2006relationship}
J.~Davis and M.~Goadrich.
\newblock The relationship between precision-recall and roc curves.
\newblock In {\em Proceedings of the International Conference on Machine
  learning (ICML)}, 2006.

\bibitem{devlin2018bert}
J.~Devlin, M.-W. Chang, K.~Lee, and K.~Toutanova.
\newblock Bert: Pre-training of deep bidirectional transformers for language
  understanding.
\newblock {\em arXiv preprint arXiv:1810.04805}, 2018.

\bibitem{hawthorne2017onsets}
C.~Hawthorne, E.~Elsen, J.~Song, A.~Roberts, I.~Simon, C.~Raffel, J.~Engel,
  S.~Oore, and D.~Eck.
\newblock Onsets and frames: Dual-objective piano transcription.
\newblock {\em arXiv preprint arXiv:1710.11153}, 2017.

\bibitem{hawthorne2018enabling}
C.~Hawthorne, A.~Stasyuk, A.~Roberts, I.~Simon, C.-Z.~A. Huang, S.~Dieleman,
  E.~Elsen, J.~Engel, and D.~Eck.
\newblock Enabling factorized piano music modeling and generation with the
  {MAESTRO} dataset.
\newblock In {\em Proceedings of the International Conference on Learning
  Representations (ICLR)}, 2019.

\bibitem{he2016deep}
K.~He, X.~Zhang, S.~Ren, and J.~Sun.
\newblock Deep residual learning for image recognition.
\newblock In {\em Proceedings of the IEEE conference on Computer Vision and
  Pattern Recognition (CVPR)}, 2016.

\bibitem{hoffer2017train}
E.~Hoffer, I.~Hubara, and D.~Soudry.
\newblock Train longer, generalize better: closing the generalization gap in
  large batch training of neural networks.
\newblock In {\em Proceedings of the Advances in Neural Information Processing
  Systems (NIPS)}, 2017.

\bibitem{huang2018musictransformer}
C.-Z.~A. Huang, A.~Vaswani, J.~Uszkoreit, I.~Simon, C.~Hawthorne, N.~Shazeer,
  A.~M. Dai, M.~D. Hoffman, M.~Dinculescu, and D.~Eck.
\newblock Music transformer.
\newblock In {\em Proceedings of the International Conference on Learning
  Representations (ICLR)}, 2019.

\bibitem{keskar2017improving}
N.~S. Keskar and R.~Socher.
\newblock Improving generalization performance by switching from adam to sgd.
\newblock {\em arXiv preprint arXiv:1712.07628}, 2017.

\bibitem{nsml2}
H.~Kim, M.~Kim, D.~Seo, J.~Kim, H.~Park, S.~Park, H.~Jo, K.~Kim, Y.~Yang,
  Y.~Kim, et~al.
\newblock {NSML}: Meet the mlaas platform with a real-world case study.
\newblock {\em arXiv preprint arXiv:1810.09957}, 2018.

\bibitem{kim2018sample}
T.~Kim, J.~Lee, and J.~Nam.
\newblock Sample-level cnn architectures for music auto-tagging using raw
  waveforms.
\newblock In {\em Proceedings of the IEEE International Conference on
  Acoustics, Speech and Signal Processing (ICASSP)}, 2018.

\bibitem{kingma2014adam}
D.~P. Kingma and J.~Ba.
\newblock Adam: A method for stochastic optimization.
\newblock In {\em Proceedings of the International Conference on Learning
  Representations (ICLR)}, 2015.

\bibitem{krebs2016downbeat}
F.~Krebs, S.~B{\"o}ck, M.~Dorfer, and G.~Widmer.
\newblock Downbeat tracking using beat synchronous features with recurrent
  neural networks.
\newblock In {\em Proceedings of the International Society for Music
  Information Retrieval Conference (ISMIR)}, 2016.

\bibitem{law2009evaluation}
E.~Law, K.~West, M.~I. Mandel, M.~Bay, and J.~S. Downie.
\newblock Evaluation of algorithms using games: The case of music tagging.
\newblock In {\em Proceedings of the International Society for Music
  Information Retrieval Conference (ISMIR)}, 2009.

\bibitem{lee2017sample}
J.~Lee, J.~Park, K.~L. Kim, and J.~Nam.
\newblock Sample-level deep convolutional neural networks for music
  auto-tagging using raw waveforms.
\newblock {\em Proceedings of the Sound and Music Computing Conference (SMC)},
  2017.

\bibitem{lee2017interactive}
J.~Lee, J.-H. Shin, and J.-S. Kim.
\newblock Interactive visualization and manipulation of attention-based neural
  machine translation.
\newblock In {\em Proceedings of the conference on Empirical Methods in Natural
  Language Processing (EMNLP): System Demonstrations}, 2017.

\bibitem{lin2017structured}
Z.~Lin, M.~Feng, C.~N.~d. Santos, M.~Yu, B.~Xiang, B.~Zhou, and Y.~Bengio.
\newblock A structured self-attentive sentence embedding.
\newblock {\em Proceedings of the International Conference on Learning
  Representations (ICLR)}, 2017.

\bibitem{mcfee2015librosa}
B.~McFee, C.~Raffel, D.~Liang, D.~P. Ellis, M.~McVicar, E.~Battenberg, and
  O.~Nieto.
\newblock librosa: Audio and music signal analysis in python.
\newblock In {\em Proceedings of the python in science conference}, 2015.

\bibitem{mishra2017local}
S.~Mishra, B.~L. Sturm, and S.~Dixon.
\newblock Local interpretable model-agnostic explanations for music content
  analysis.
\newblock In {\em Proceedings of the International Society for Music
  Infor-mation Retrieval Conference (ISMIR)}, 2017.

\bibitem{mishraunderstanding}
S.~Mishra, B.~L. Sturm, and S.~Dixon.
\newblock Understanding a deep machine listening model through feature
  inversion.
\newblock In {\em Proceedings of the International Society for Music
  Infor-mation Retrieval Conference (ISMIR)}, 2018.

\bibitem{parmar2018imagetransformer}
N.~Parmar, A.~Vaswani, J.~Uszkoreit, {\L}.~Kaiser, N.~Shazeer, A.~Ku, and
  D.~Tran.
\newblock Image transformer.
\newblock {\em Proceedings of the International Conference on Machine learning
  (ICML)}, 2018.

\bibitem{pascanu2013difficulty}
R.~Pascanu, T.~Mikolov, and Y.~Bengio.
\newblock On the difficulty of training recurrent neural networks.
\newblock In {\em Proceedings of the International Conference on Machine
  learning (ICML)}, 2013.

\bibitem{pons2017end}
J.~Pons, O.~Nieto, M.~Prockup, E.~Schmidt, A.~Ehmann, and X.~Serra.
\newblock End-to-end learning for music audio tagging at scale.
\newblock {\em Proceedings of the International Society for Music Information
  Retrieval Conference (ISMIR)}, 2018.

\bibitem{pons2017designing}
J.~Pons and X.~Serra.
\newblock Designing efficient architectures for modeling temporal features with
  convolutional neural networks.
\newblock In {\em Proceedings of the IEEE International Conference on
  Acoustics, Speech and Signal Processing (ICASSP)}, 2017.

\bibitem{pons2017timbre}
J.~Pons, O.~Slizovskaia, R.~Gong, E.~G{\'o}mez, and X.~Serra.
\newblock Timbre analysis of music audio signals with convolutional neural
  networks.
\newblock In {\em Proceedings of the European Signal Processing Conference
  (EUSIPCO)}, 2017.

\bibitem{radford2018language}
A.~Radford, J.~Wu, R.~Child, D.~Luan, D.~Amodei, and I.~Sutskever.
\newblock Language models are unsupervised multitask learners.
\newblock Technical report, Technical report, OpenAi, 2018.

\bibitem{roberts2017hierarchical}
A.~Roberts, J.~Engel, and D.~Eck.
\newblock Hierarchical variational autoencoders for music.
\newblock In {\em NIPS Workshop on Machine Learning for Creativity and Design},
  2017.

\bibitem{seong2018towards}
S.~Seong, Y.~Lee, Y.~Kee, D.~Han, and J.~Kim.
\newblock Towards flatter loss surface via nonmonotonic learning rate
  scheduling.
\newblock In {\em Conference on Uncertainty in Artificial Intelligence (UAI)},
  2018.

\bibitem{sigtia2016end}
S.~Sigtia, E.~Benetos, and S.~Dixon.
\newblock An end-to-end neural network for polyphonic piano music
  transcription.
\newblock {\em IEEE/ACM Transactions on Audio, Speech, and Language
  Processing}, 2016.

\bibitem{simonyan2014very}
K.~Simonyan and A.~Zisserman.
\newblock Very deep convolutional networks for large-scale image recognition.
\newblock {\em Proceedings of the International Conference on Learning
  Representations (ICLR)}, 2015.

\bibitem{nsml1}
N.~Sung, M.~Kim, H.~Jo, Y.~Yang, J.~Kim, L.~Lausen, Y.~Kim, G.~Lee, D.~Kwak,
  J.-W. Ha, et~al.
\newblock {NSML}: A machine learning platform that enables you to focus on your
  models.
\newblock {\em arXiv preprint arXiv:1712.05902}, 2017.

\bibitem{tang2015document}
D.~Tang, B.~Qin, and T.~Liu.
\newblock Document modeling with gated recurrent neural network for sentiment
  classification.
\newblock In {\em Proceedings of the conference on Empirical Methods in Natural
  Language Processing (EMNLP)}, 2015.

\bibitem{van2016wavenet}
A.~Van Den~Oord, S.~Dieleman, H.~Zen, K.~Simonyan, O.~Vinyals, A.~Graves,
  N.~Kalchbrenner, A.~W. Senior, and K.~Kavukcuoglu.
\newblock Wavenet: A generative model for raw audio.
\newblock {\em Speech Synthesis Workshop (SSW)}, 2016.

\bibitem{vaswani2017attention}
A.~Vaswani, N.~Shazeer, N.~Parmar, J.~Uszkoreit, L.~Jones, A.~N. Gomez,
  {\L}.~Kaiser, and I.~Polosukhin.
\newblock Attention is all you need.
\newblock In {\em Proceedings of the Advances in Neural Information Processing
  Systems (NIPS)}, pages 5998--6008, 2017.

\bibitem{vogl2017drum}
R.~Vogl, M.~Dorfer, G.~Widmer, and P.~Knees.
\newblock Drum transcription via joint beat and drum modeling using
  convolutional recurrent neural networks.
\newblock In {\em Proceedings of the International Society for Music
  Information Retrieval Conference (ISMIR)}, 2017.

\bibitem{wang2018nonlocal}
X.~Wang, R.~Girshick, A.~Gupta, and K.~He.
\newblock Non-local neural networks.
\newblock In {\em Proceedings of the IEEE conference on Computer Vision and
  Pattern Recognition (CVPR)}, 2018.

\bibitem{wilson2017marginal}
A.~C. Wilson, R.~Roelofs, M.~Stern, N.~Srebro, and B.~Recht.
\newblock The marginal value of adaptive gradient methods in machine learning.
\newblock In {\em Proceedings of the Advances in Neural Information Processing
  Systems (NIPS)}, 2017.

\bibitem{DBLP:journals/corr/abs-1810-13125}
K.~Yue, M.~Sun, Y.~Yuan, F.~Zhou, E.~Ding, and F.~Xu.
\newblock Compact generalized non-local network.
\newblock {\em Proceedings of the Advances in Neural Information Processing
  Systems (NIPS)}, 2018.

\bibitem{zhang2018sagan}
H.~Zhang, I.~Goodfellow, D.~Metaxas, and A.~Odena.
\newblock Self-attention generative adversarial networks.
\newblock {\em Proceedings of the International Conference on Machine learning
  (ICML)}, 2019.

\bibitem{zuo2015convolutional}
Z.~Zuo, B.~Shuai, G.~Wang, X.~Liu, X.~Wang, B.~Wang, and Y.~Chen.
\newblock Convolutional recurrent neural networks: Learning spatial
  dependencies for image representation.
\newblock In {\em Proceedings of the IEEE conference on Computer Vision and
  Pattern Recognition (CVPR)}, 2015.

\end{thebibliography}
}

\newpage
\onecolumn
\appendix

\section{Per tag AUROC on MTAT}

In Table~\ref{table:pertagauroc}, we report per tag AUROC on MTAT in a descending order. Note that our model is vulnerable to negative tags: `no voice', `no vocal', and `no vocals'.

\begin{table}[h!]
\centering
\small
\begin{tabular}{@{}ccccc@{}}
\toprule
metal        & choral & choir  & rock       & opera   \\
98.79        & 98.68  & 98.65  & 98.52      & 98.34   \\\midrule
flute      & harpsichord & cello        & techno    & dance    \\
97.91      & 97.85       & 96.61        & 96.51     & 96.08    \\\midrule
ambient      & piano  & harp   & country    & pop \\
95.69        & 95.69  & 95.12  & 94.05      & 94.01  \\\midrule
sitar      & man         & woman        & female    & beat     \\
93.99      & 93.70       & 93.53        & 93.41     & 93.23    \\\midrule
female vocal & male   & violin & male vocal & beats \\
92.99        & 92.91  & 92.62  & 92.34      & 92.25  \\\midrule
classical  & loud        & female voice & guitar    & quiet    \\
92.15      & 91.30       & 91.05        & 90.98     & 90.89    \\\midrule
solo         & drums  & indian & male voice & singing \\
90.36        & 89.99  & 89.99  & 89.79      & 89.77  \\\midrule
electronic & fast        & vocal        & new age   & classic  \\
89.37      & 89.15       & 88.81        & 88.73     & 88.66    \\\midrule
strings      & vocals & synth  & voice      & slow   \\
88.59        & 88.11  & 86.17  & 84.60      & 84.09   \\\midrule
soft       & weird       & no vocal     & no vocals & no voice \\
83.63      & 81.68       & 71.86        & 70.22     & 67.88    \\ \bottomrule
\end{tabular}
\vspace{0.3cm}
\caption{Per tag AUROC on MTAT}
\vspace{-0.5cm}
\label{table:pertagauroc}
\end{table}

\section{More Results on Attention Heat Maps}

\begin{figure*}[ht!]
    \centering
    \begin{subfigure}[h]{0.48\linewidth}
        \centering
        \includegraphics[width=\linewidth]{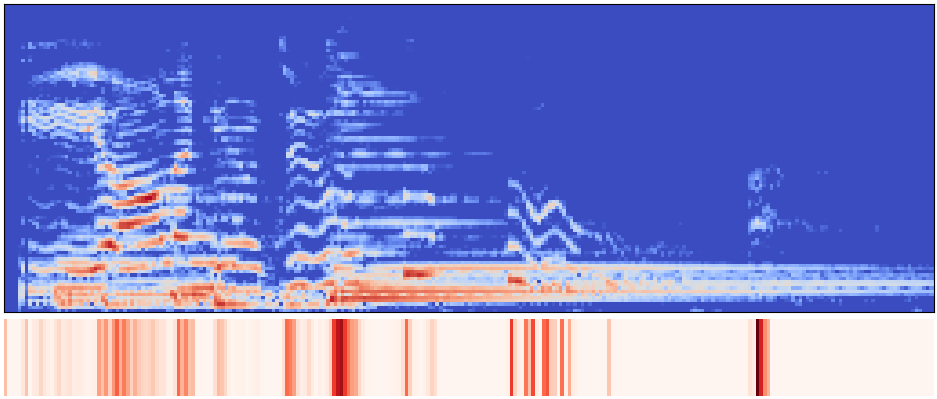}
        \caption{Tag - Male}
    \end{subfigure}
    \begin{subfigure}[h]{0.48\linewidth}
        \centering
        \includegraphics[width=\linewidth]{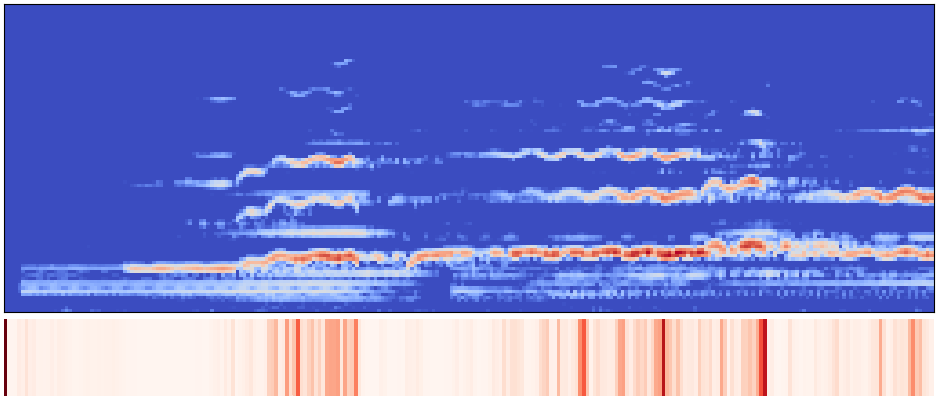}
        \caption{Tag - Female}
    \end{subfigure}
    \begin{subfigure}[h]{0.48\linewidth}
        \centering
        \includegraphics[width=\linewidth]{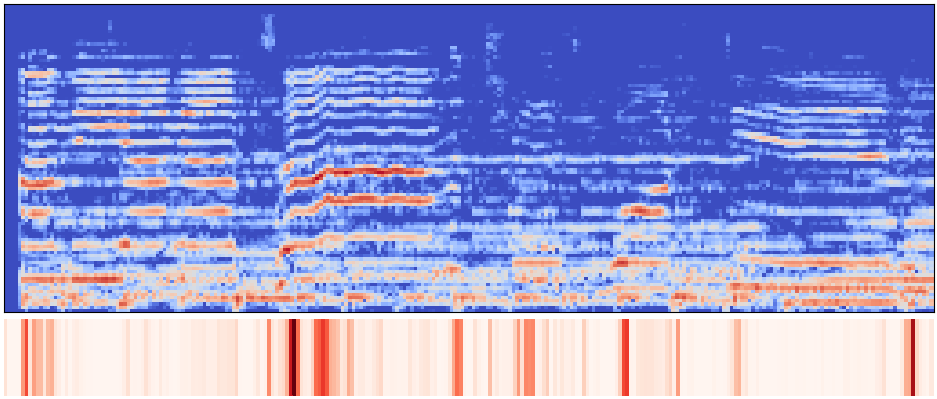}
        \caption{Tag - Vocal}
    \end{subfigure}
    \begin{subfigure}[h]{0.48\linewidth}
        \centering
        \includegraphics[width=\linewidth]{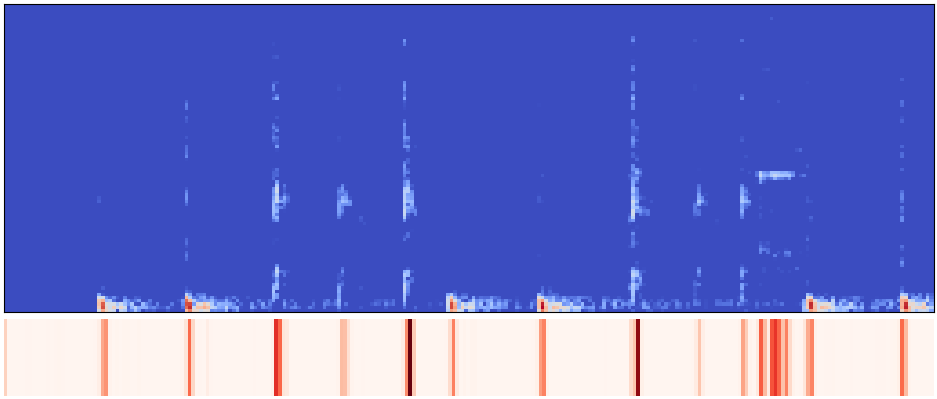}
        \caption{Tag - No Vocal}
    \end{subfigure}
    \caption{Attention heat maps for voice tags.}
    \label{fig:attention_heatmap_voice}
    \vspace{-0.5cm}
\end{figure*}

\begin{figure*}[ht!]
    \centering
    \begin{subfigure}[h]{0.48\linewidth}
        \centering
        \includegraphics[width=\linewidth]{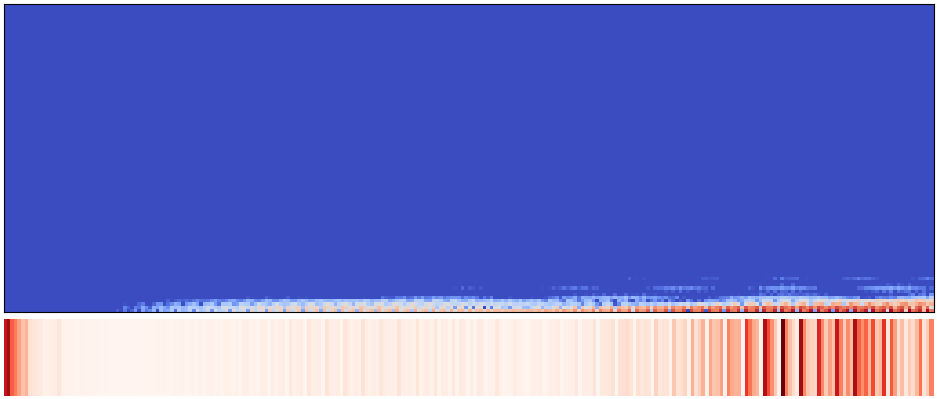}
        \caption{Tag - Quiet}
    \end{subfigure}
    \begin{subfigure}[h]{0.48\linewidth}
        \centering
        \includegraphics[width=\linewidth]{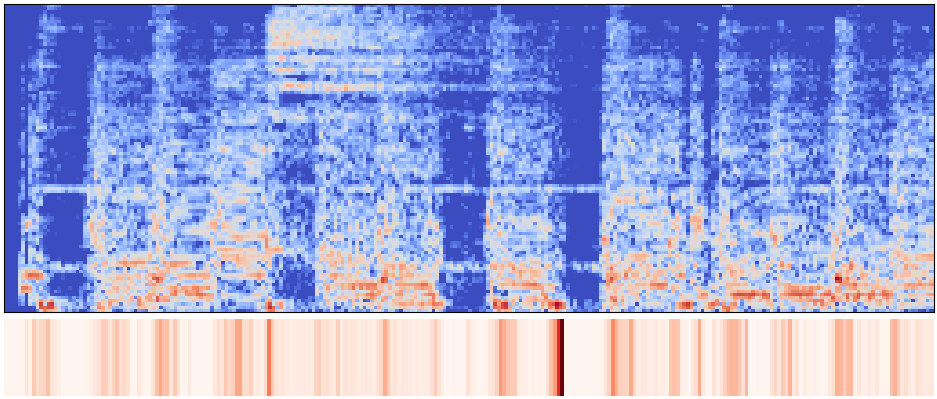}
        \caption{Tag - Loud}
    \end{subfigure}
    \begin{subfigure}[h]{0.48\linewidth}
        \centering
        \includegraphics[width=\linewidth]{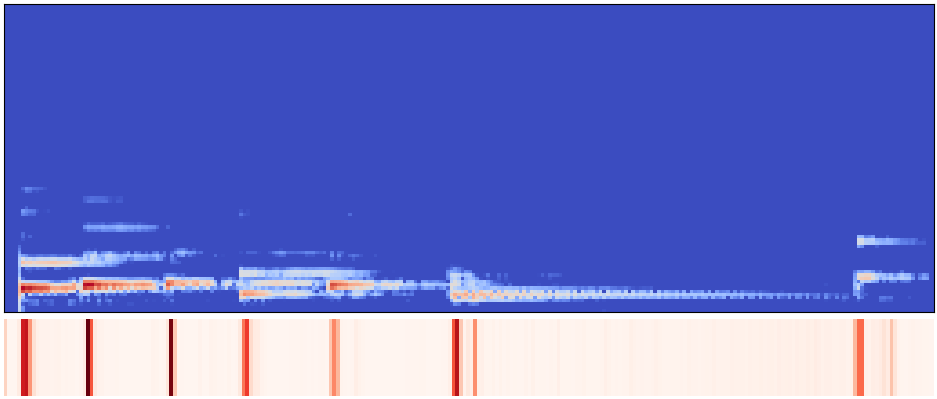}
        \caption{Tag - Slow}
    \end{subfigure}
    \begin{subfigure}[h]{0.48\linewidth}
        \centering
        \includegraphics[width=\linewidth]{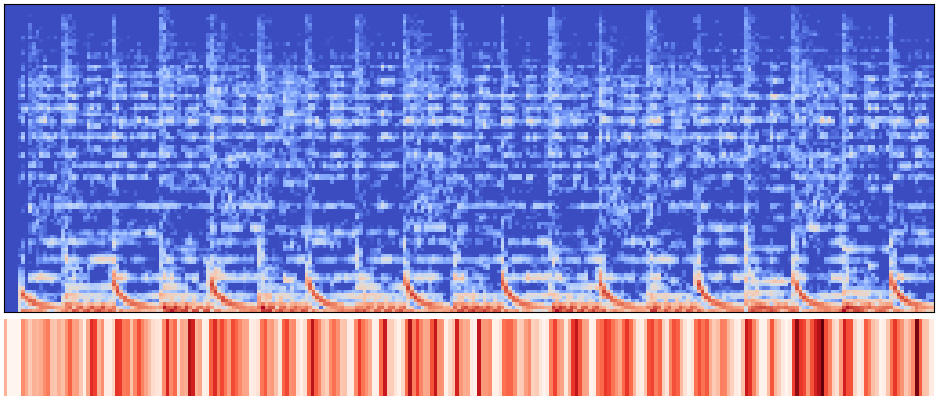}
        \caption{Tag - Fast}
    \end{subfigure}
    \begin{subfigure}[h]{0.48\linewidth}
        \centering
        \includegraphics[width=\linewidth]{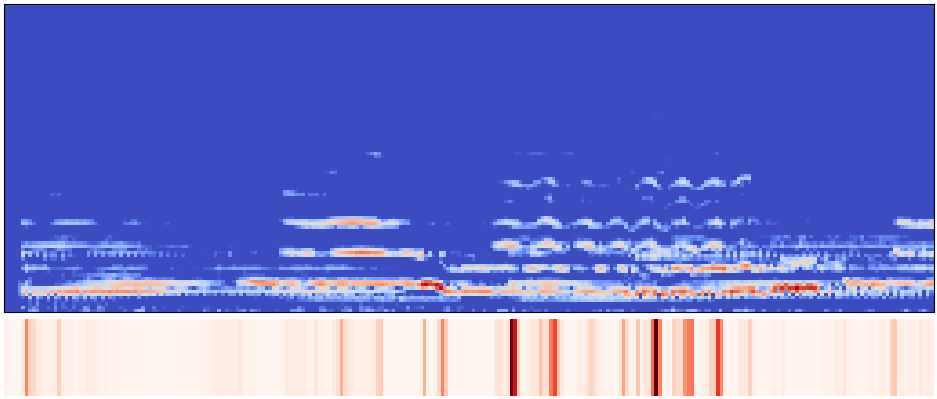}
        \caption{Tag - Soft}
    \end{subfigure}
    \begin{subfigure}[h]{0.48\linewidth}
        \centering
        \includegraphics[width=\linewidth]{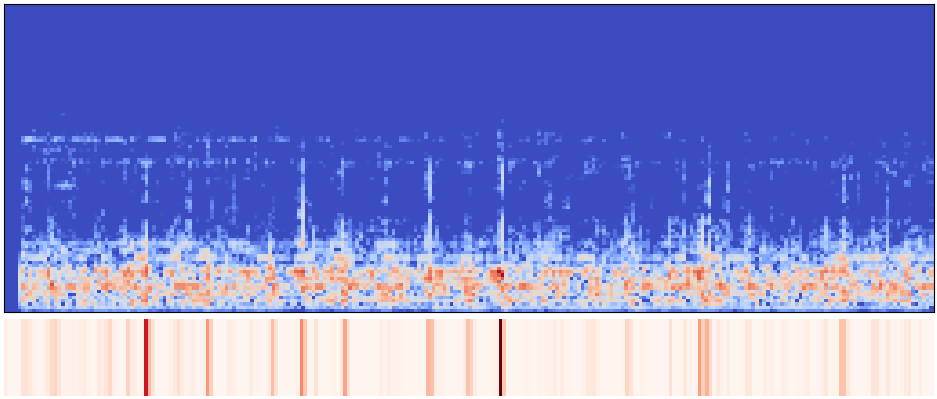}
        \caption{Tag - Weird}
    \end{subfigure}
    \caption{Attention heat maps for mood tags.}
    \label{fig:attention_heatmap_mood}
\end{figure*}

\begin{figure*}[h!]
    \centering
    \begin{subfigure}[h]{0.48\linewidth}
        \centering
        \includegraphics[width=\linewidth]{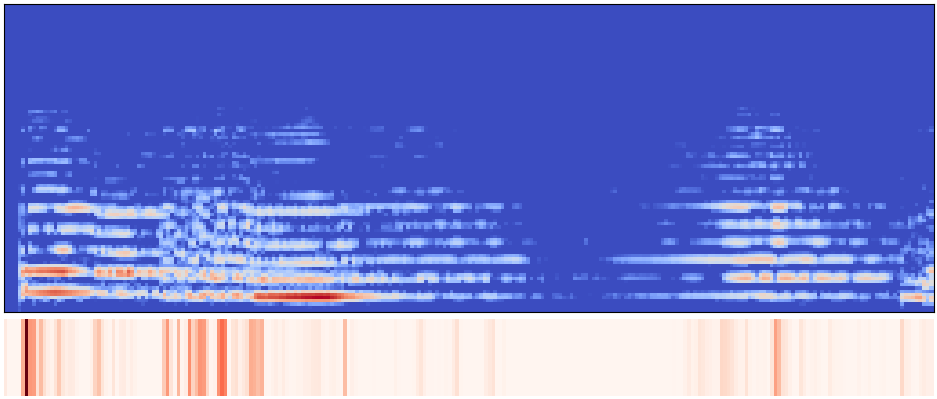}
        \caption{Tag - Cello}
    \end{subfigure}
    \begin{subfigure}[h]{0.48\linewidth}
        \centering
        \includegraphics[width=\linewidth]{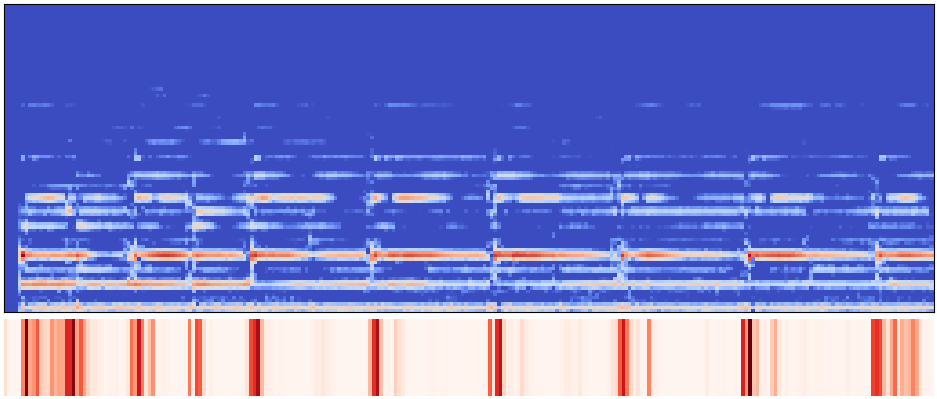}
        \caption{Tag - Sitar}
    \end{subfigure}
    \begin{subfigure}[h]{0.48\linewidth}
        \centering
        \includegraphics[width=\linewidth]{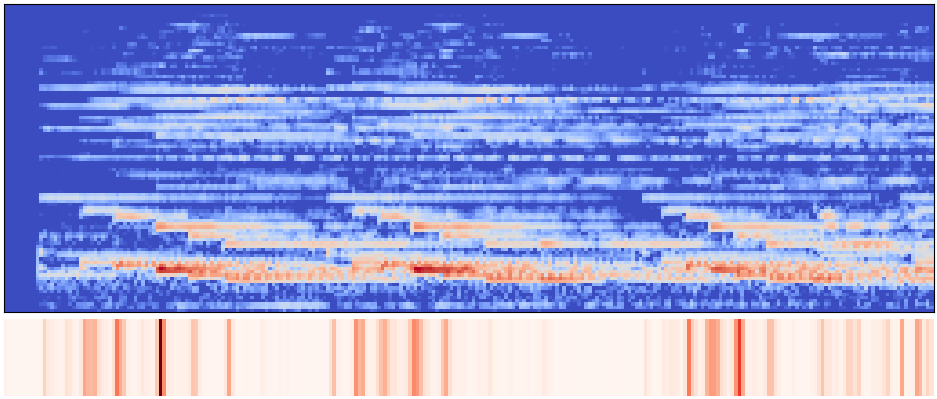}
        \caption{Tag - Harp}
    \end{subfigure}
    \begin{subfigure}[h]{0.48\linewidth}
        \centering
        \includegraphics[width=\linewidth]{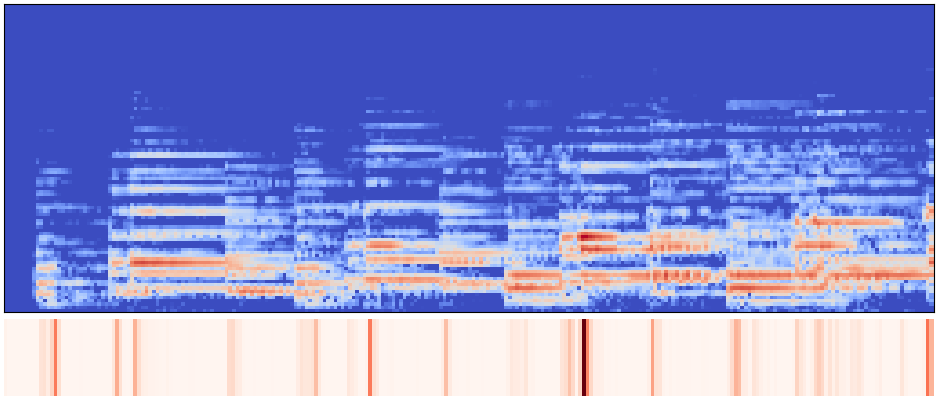}
        \caption{Tag - Piano}
    \end{subfigure}
    \begin{subfigure}[h]{0.48\linewidth}
        \centering
        \includegraphics[width=\linewidth]{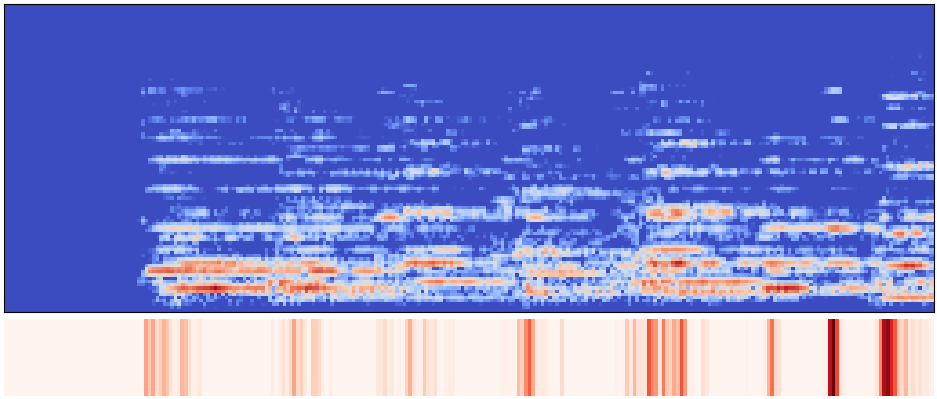}
        \caption{Tag - Strings}
    \end{subfigure}
    \begin{subfigure}[h]{0.48\linewidth}
        \centering
        \includegraphics[width=\linewidth]{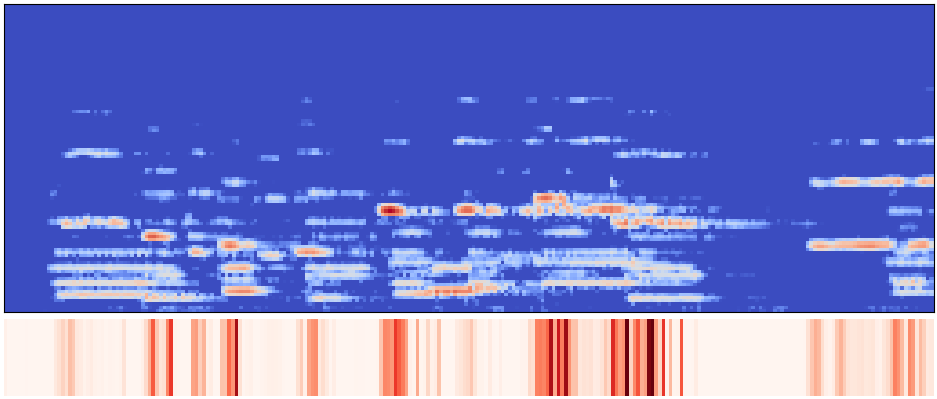}
        \caption{Tag - Flute}
    \end{subfigure}
    \begin{subfigure}[h]{0.48\linewidth}
        \centering
        \includegraphics[width=\linewidth]{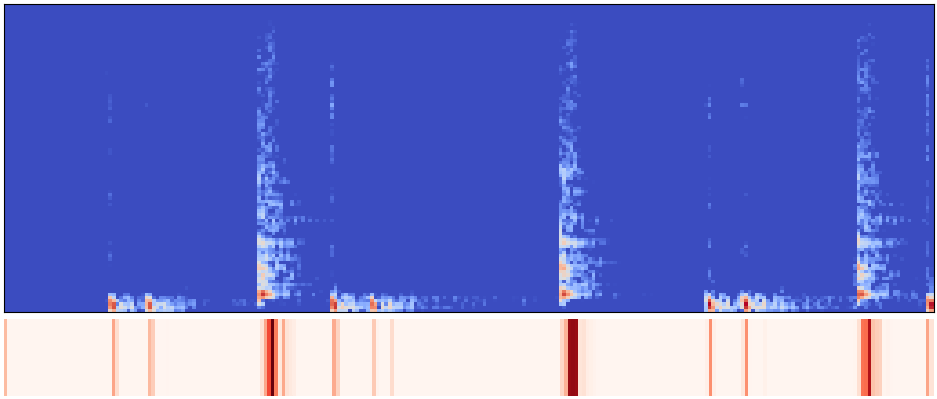}
        \caption{Tag - Drums}
    \end{subfigure}
    \begin{subfigure}[h]{0.48\linewidth}
        \centering
        \includegraphics[width=\linewidth]{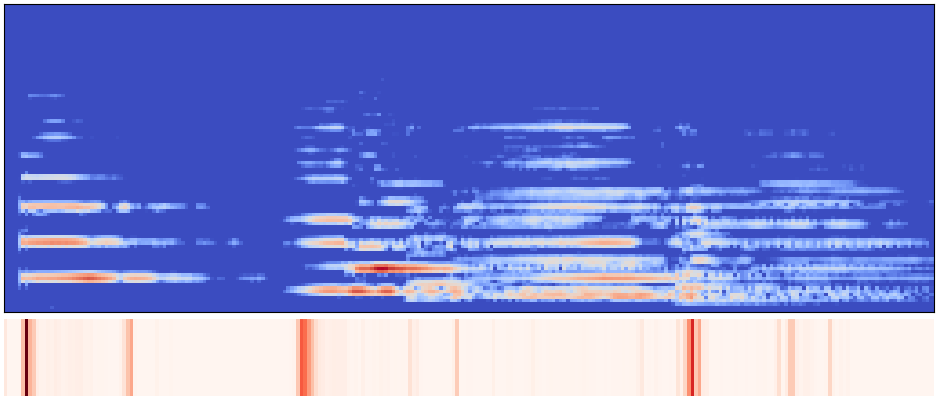}
        \caption{Tag - Violin}
    \end{subfigure}
    \begin{subfigure}[h]{0.48\linewidth}
        \centering
        \includegraphics[width=\linewidth]{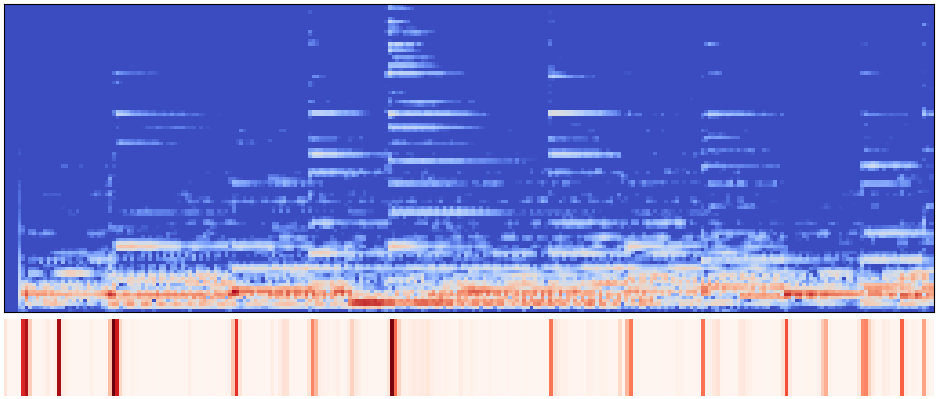}
        \caption{Tag - Guitar}
    \end{subfigure}
    \begin{subfigure}[h]{0.48\linewidth}
        \centering
        \includegraphics[width=\linewidth]{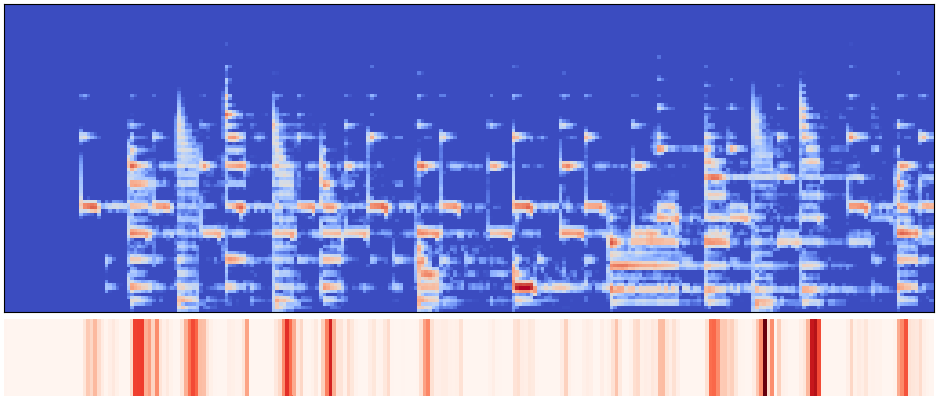}
        \caption{Tag - Synth}
    \end{subfigure}
    \caption{Attention heat maps for instrument tags.}
    \label{fig:attention_heatmap_inst}
\end{figure*}

\begin{figure*}[h!]
    \centering
    \begin{subfigure}[h]{0.48\linewidth}
        \centering
        \includegraphics[width=\linewidth]{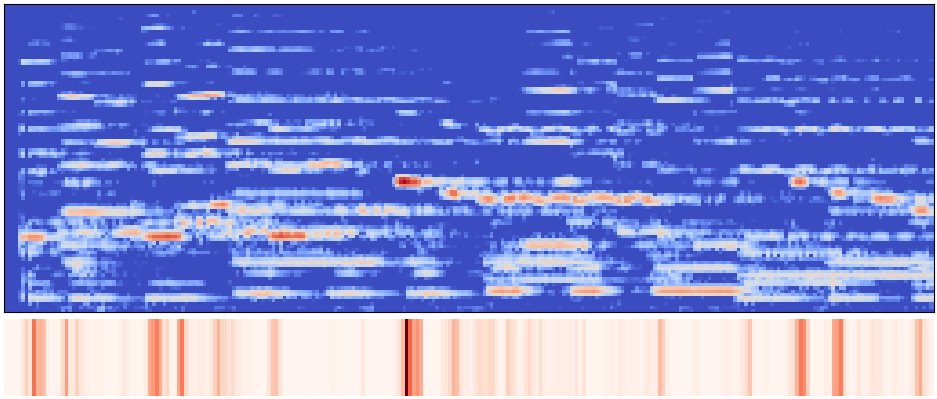}
        \caption{Tag - Classic}
    \end{subfigure}
    \begin{subfigure}[h]{0.48\linewidth}
        \centering
        \includegraphics[width=\linewidth]{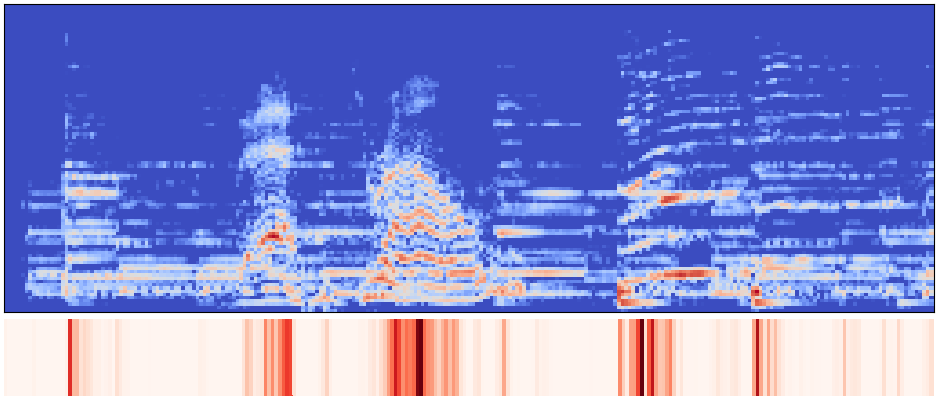}
        \caption{Tag - Country}
    \end{subfigure}
    \begin{subfigure}[h]{0.48\linewidth}
        \centering
        \includegraphics[width=\linewidth]{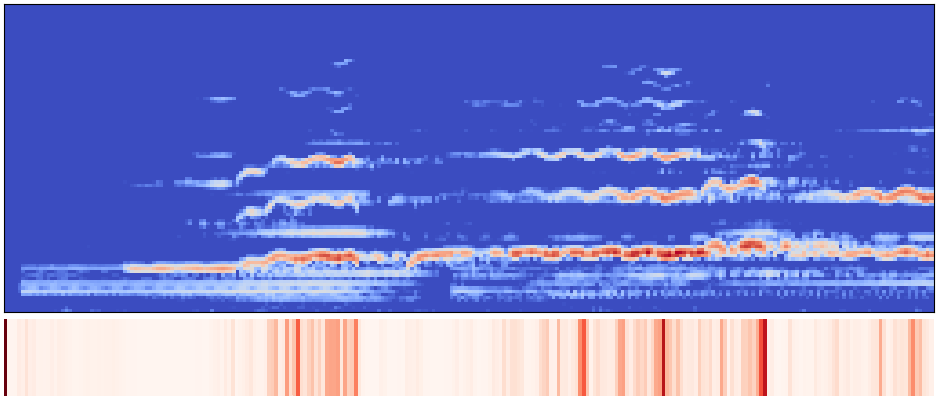}
        \caption{Tag - Opera}
    \end{subfigure}
    \begin{subfigure}[h]{0.48\linewidth}
        \centering
        \includegraphics[width=\linewidth]{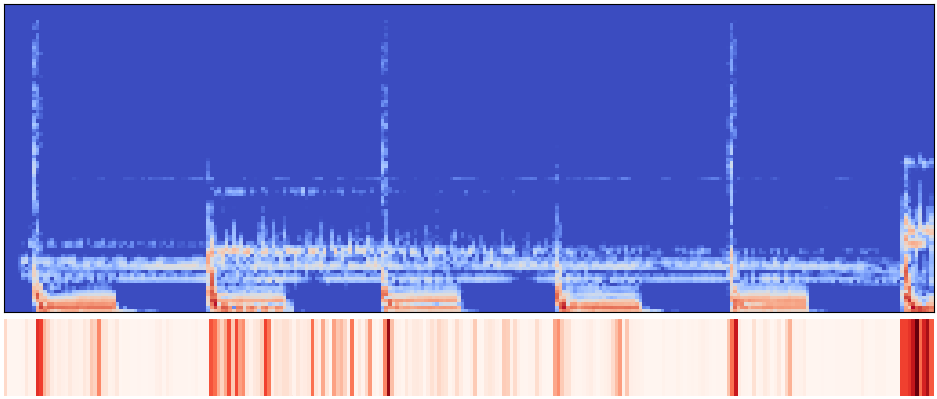}
        \caption{Tag - New Age}
    \end{subfigure}
    \begin{subfigure}[h]{0.48\linewidth}
        \centering
        \includegraphics[width=\linewidth]{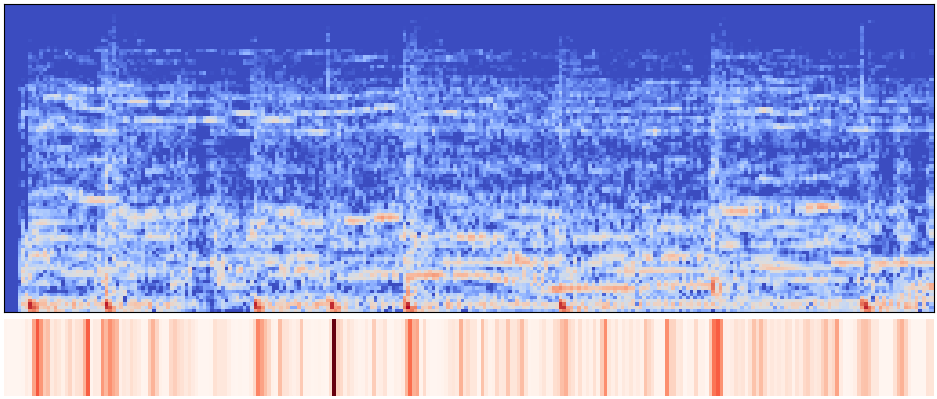}
        \caption{Tag - Rock}
    \end{subfigure}
    \begin{subfigure}[h]{0.48\linewidth}
        \centering
        \includegraphics[width=\linewidth]{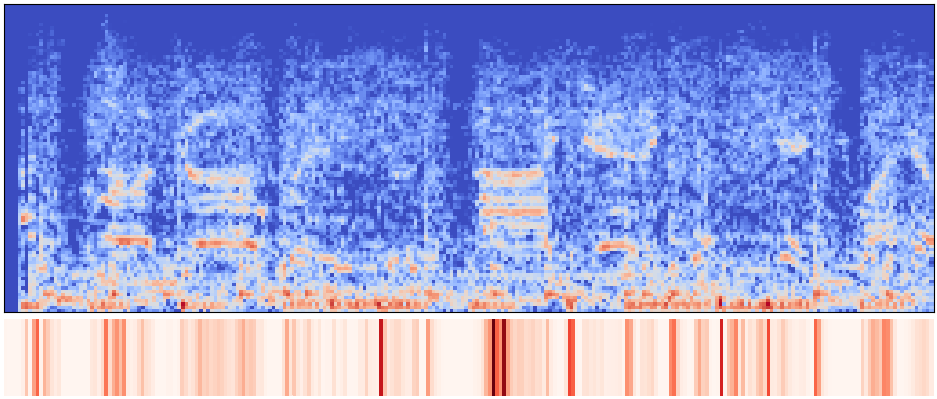}
        \caption{Tag - Metal}
    \end{subfigure}
    \begin{subfigure}[h]{0.48\linewidth}
        \centering
        \includegraphics[width=\linewidth]{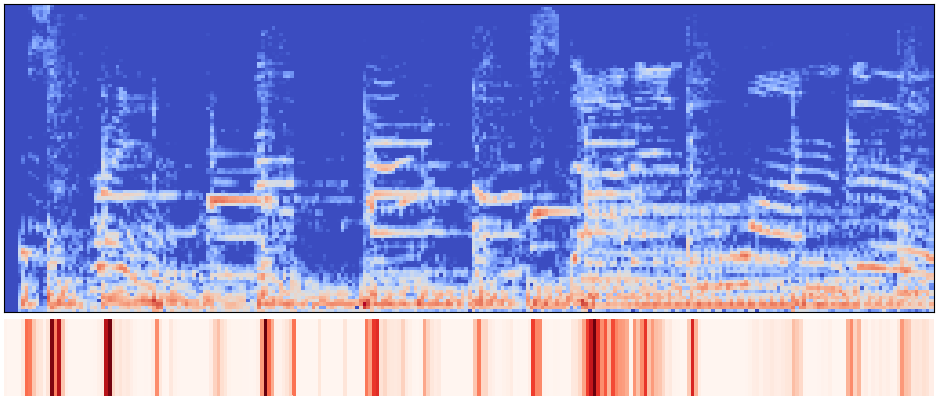}
        \caption{Tag - Pop}
    \end{subfigure}
    \begin{subfigure}[h]{0.48\linewidth}
        \centering
        \includegraphics[width=\linewidth]{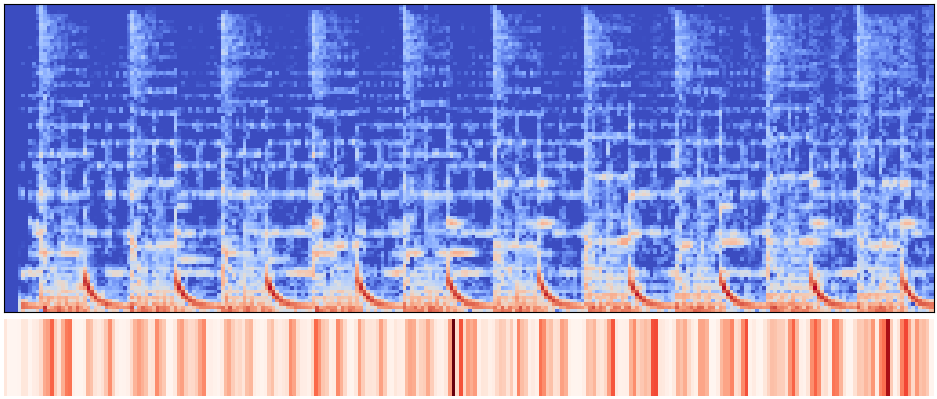}
        \caption{Tag - Dance}
    \end{subfigure}
    \begin{subfigure}[h]{0.48\linewidth}
        \centering
        \includegraphics[width=\linewidth]{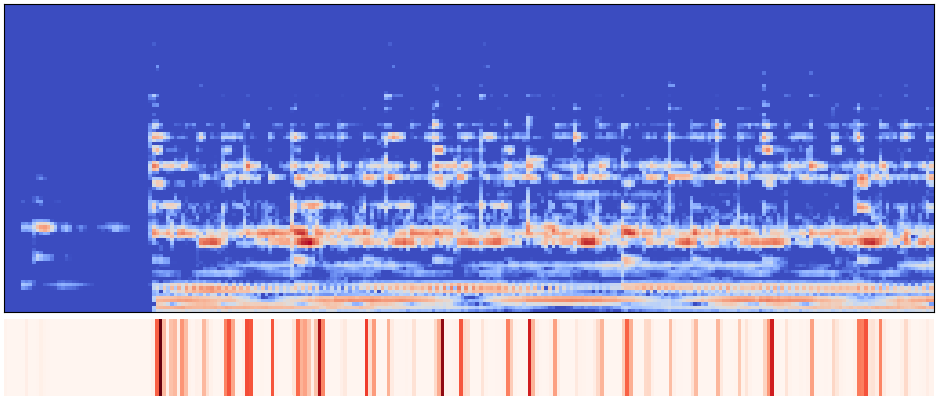}
        \caption{Tag - Electronic}
    \end{subfigure}
    \begin{subfigure}[h]{0.48\linewidth}
        \centering
        \includegraphics[width=\linewidth]{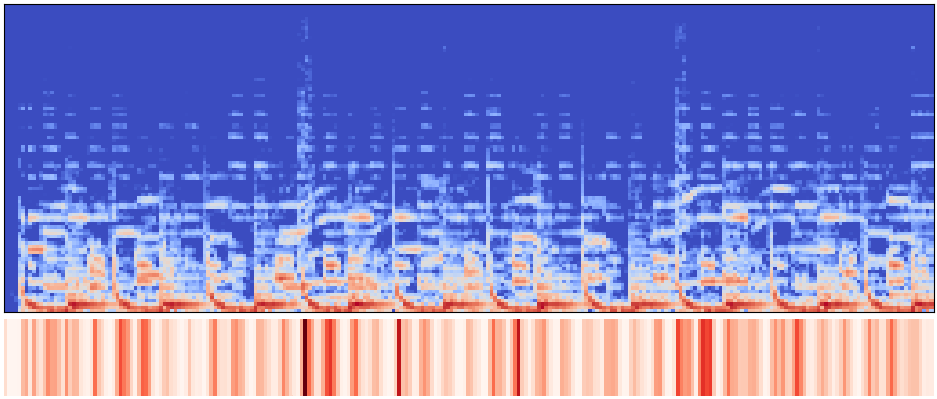}
        \caption{Tag - Techno}
    \end{subfigure}
    \caption{Attention heat maps for genre tags.}
    \label{fig:attention_heatmap_gnr}
\end{figure*}

We report more attention heat maps of various types including voice (Figure~\ref{fig:attention_heatmap_voice}), mood (Figure~\ref{fig:attention_heatmap_mood}),  instrument (Figure~\ref{fig:attention_heatmap_inst}), and genre (Figure~\ref{fig:attention_heatmap_gnr}).

\section{Tag-wise Contribution Heat Maps}

\begin{figure*}[ht!]
    \centering
    \begin{subfigure}[h]{0.48\linewidth}
        \includegraphics[width=\linewidth]{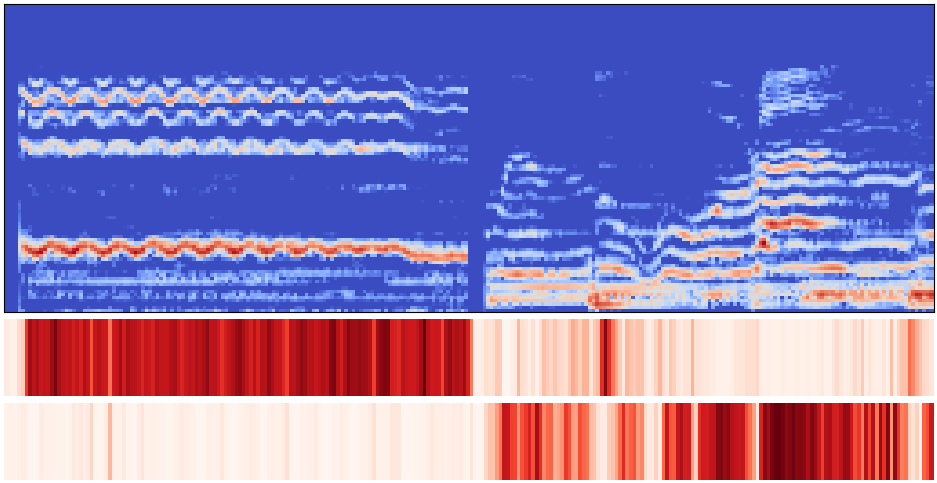}
        \caption{Female + Male}
    \end{subfigure}
    \begin{subfigure}[h]{0.48\linewidth}
        \includegraphics[width=\linewidth]{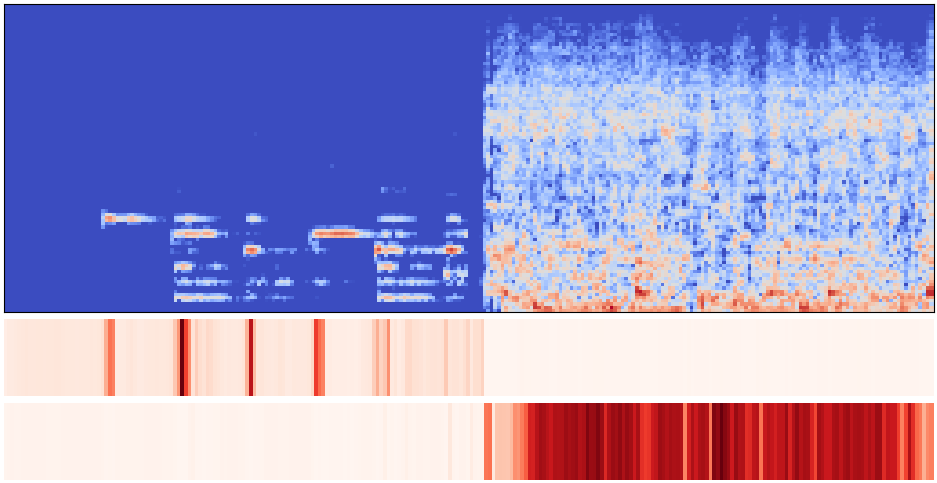}
        \caption{Classic + Metal}
    \end{subfigure}
    \begin{subfigure}[h]{0.48\linewidth}
        \includegraphics[width=\linewidth]{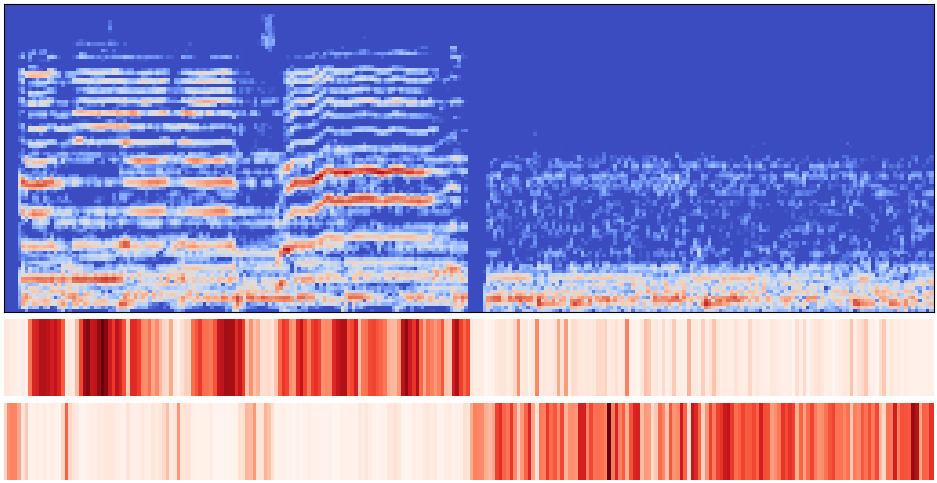}
        \caption{Vocal + No Vocals}
    \end{subfigure}
    \begin{subfigure}[h]{0.48\linewidth}
        \includegraphics[width=\linewidth]{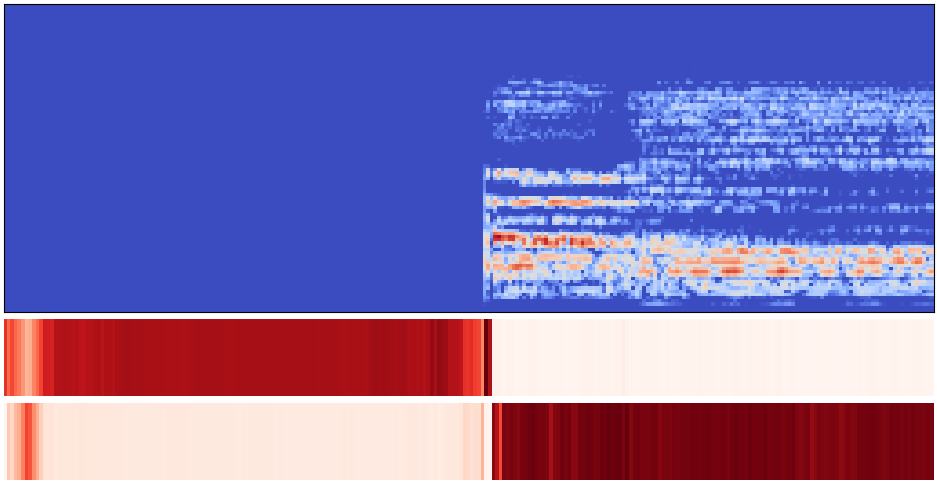}
        \caption{No Voice + Choir}
    \end{subfigure}
    \begin{subfigure}[h]{0.48\linewidth}
        \includegraphics[width=\linewidth]{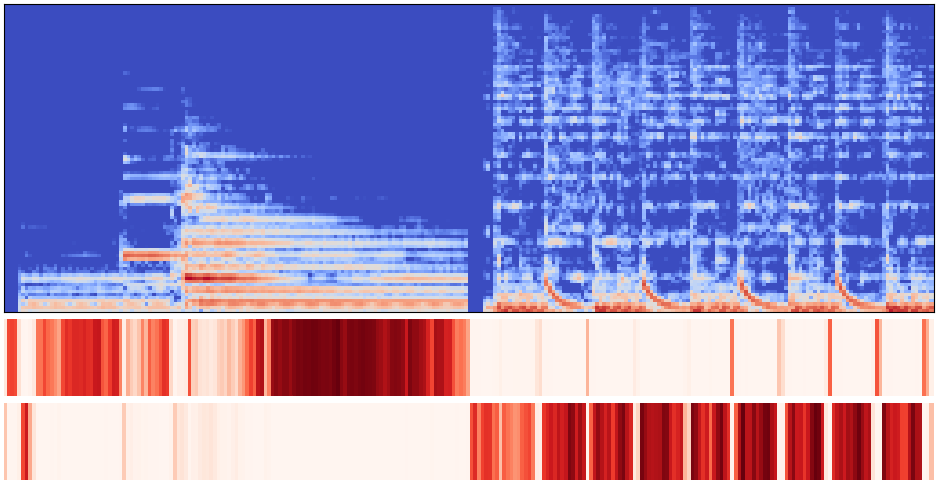}
        \caption{Slow + Fast}
    \end{subfigure}
    \begin{subfigure}[h]{0.48\linewidth}
        \includegraphics[width=\linewidth]{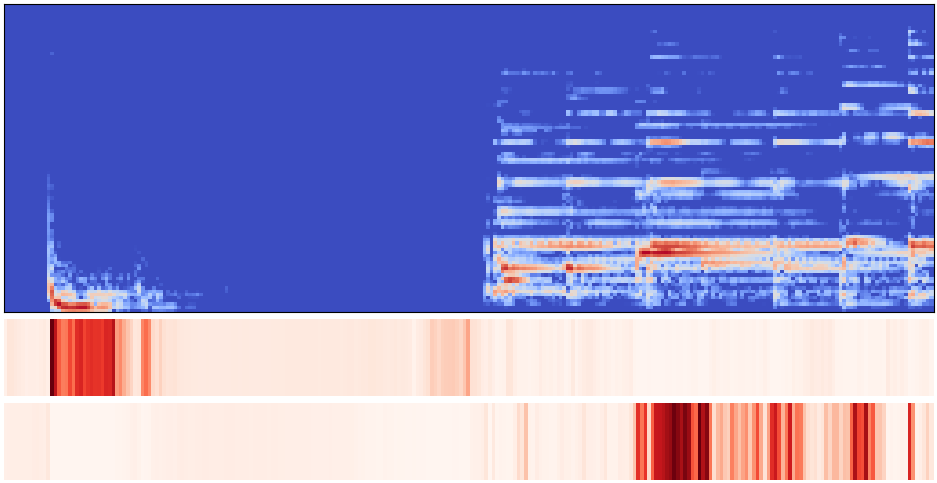}
        \caption{Drums + Harp}
    \end{subfigure}
    \caption{Tag-wise contribution heat maps.}
    \label{fig:tagwise_contrib_sub}
\end{figure*}

More tag-wise contribution heat maps are illustrated in Figure~\ref{fig:tagwise_contrib_sub}.

\newpage

\end{document}